\documentclass[12pt]{article}

\usepackage[utf8]{inputenc}
\usepackage[english]{babel} 
\usepackage[T1]{fontenc} 

\usepackage{authblk}  

\makeatletter 
\renewcommand\AB@affilnote[1]{\textsuperscript{\scriptsize #1}}
\makeatother

\usepackage{amsmath,amssymb,mathtools,amsthm}
\allowdisplaybreaks[1]

\usepackage{amsfonts}
\usepackage{graphicx}
\usepackage{enumerate}
\usepackage{enumitem}
\usepackage{hyperref}
\usepackage{cleveref}
\usepackage{multirow}
\usepackage{array}  
\usepackage{caption}
\usepackage{subcaption}
\usepackage[table,xcdraw,dvipsnames]{xcolor}
\definecolor{DarkGray}{HTML}{778190}   
\usepackage{arydshln} 
\usepackage{booktabs} 
\usepackage{geometry} 
\usepackage{float}

\newtheorem{theo}{Theorem}[section]

\newtheorem{rem}[theo]{Remark}
\newtheorem{example}[theo]{Example}

\begin{document}
\title{Enriching the Felsenthal index with a priori unions for decision-making processes}
\author[1]{A. Mascareñas-Pazos * }
\author[1]{S. Lorenzo-Freire}
\author[2]{J. M. Alonso-Meijide}

\affil[1]{ \emph{MODES Research Group, Department of Mathematics, Faculty of Computer Science and CITIC, University of A Coruña, Campus de Elviña, 15071 A Coruña and CITIC, 15008 A Coruña, Spain.}}
\affil[2]{ \emph{MODESTYA Research Group, Department of Statistics, Mathematical Analysis and Optimization, Faculty of Sciences, University of Santiago de Compostela, Campus de Lugo, 27002 Lugo and CITMAga, 15782 Santiago de Compostela, Spain.}}
\date{}
\maketitle
\section*{Abstract}
Within the domain of game theory, power indexes are defined as functions that quantify the influence of individual participants in collective decision-making processes. Felsenthal [D. Felsenthal. A Well-Behaved Index of a Priori P-Power for Simple N-Person
Games. \textit{Homo Oeconomicus}, 33, 2016] proposed a power index with a focus on least size winning coalitions, i.e., those coalitions capable of determining the final outcome and with the smallest number of players among all winning coalitions. However, the Felsenthal index overlooks pre-existing affinities between the players, a common and impactful factor in real-world political and economic contexts.
This paper introduces the \textit{Felsenthal Owen power index}, a novel index based on Felsenthal’s approach that integrates player affinities using Owen's a priori unions framework. The new index is rigorously 
 characterised by two distinct sets of axiomatic properties. We demonstrate its practical utility by applying it to the International Monetary Fund's voting system, revealing how strategic alliances significantly reshape power distributions.  The index thus offers policymakers a more sophisticated tool for measuring influence in complex decision-making scenarios.
 
 \vspace{1mm}
 \noindent\textbf{Keywords:} Simple games, Power, Felsenthal index, Coalitions, International Monetary Fund.

\renewcommand{\thefootnote}{}
\footnotetext{{\scriptsize  * Corresponding author.
E-mail adresses:  \texttt{alicia.mascarenas@udc.es} (A. Mascareñas-Pazos),
\texttt{silvia.lorenzo@udc.es} (S. Lorenzo-Freire), and \texttt{josemaria.alonso@usc.es} (J.M. Alonso-Meijide).
}}
\addtocounter{footnote}{-1}
\renewcommand{\thefootnote}{\arabic{footnote}}
\setcounter{footnote}{0}

\section{Introduction}
In scenarios where a collective body needs to make a decision, and its members may hold differing opinions, voting is typically employed as a method to reach a verdict. Often, voting systems are designed with assigned weights to grant some voters greater influence than others. However, using weights to reflect the power of different participants can result in misleading conclusions. A simple, yet illustrative example is a 3-member vote with appointed weights of 4,2, and 1, and where the “winning” option is awarded by a simple majority (4 out of 7). In this situation, the player with greatest weight clearly holds all the power. The other two participants are powerless despite possessing 3 of the total votes,  meaning their power doesn’t align with their relative assigned weights. Therefore, it is essential to develop alternative strategies other than weight that provide reliable insights into the true influence of participants in decision-making contexts. This is particularly important in areas such as designing fair democratic systems or assessing strategic positions in bargaining, among others.

This challenge can be addressed within the framework of cooperative game theory, where voting scenarios are mathematically modelled as simple games, and the voting influence of each player is measured  using ``power indices”.


Two of the best well-known power indices in this context are the Shapley-Shubik index~\cite{shapley_shubik} and the normalized Banzhaf index~\cite{banzhaf_power_index}.
They are both based on the relative frequency in which a player can play a pivotal role,  meaning the player can change the outcome of a coalition from losing to winning by joining it. 
Building upon these early measures, Deegan and Packel introduced in \cite{deegan_packel_power_index} an alternative index that considered only the probability of belonging to the set of minimal winning coalitions- those where every member is necessary to secure a win. More recently,  Felsenthal proposed a novel perspective in \cite{7_felsenthal_power_index} by developing an index that focused exclusively on least sized minimal winning coalitions, rather than on the entire set.  His approach is based on the idea that, in contexts like government formation, participants seek not only to be part of a winning coalition but also to maximize their individual power.
 Based on this reasoning, the Felsenthal index operates on the assumption that only winning coalitions of least size   will form, as this ensures each member receives the largest possible share of the rewards.

Up to this point, all the mentioned indexes  disregard the specific identities or attributes of individual players. In 1977, Owen defined a sophistication of the Shapley-Shubik index~\cite{25_owen_value} that accounted for pre-existing relationships between the players, such as  external agreements, affinities, organizational structures, or predefined rules. His approach was to divide the set of players into groups called ``a priori unions'', a coalition structure that is predefined before the game or  decision-making process begins. 
Building on Owen's idea,  other preexisting indexes have evolved to incorporate unions and are now referred to as coalitional power indexes. For example, the Banzhaf-Owen value~\cite{banzhaf_uniones} or the Deegan–Packel index for simple games with a priori unions~\cite{4_deegan_packel_uniones} are generalizations of the previous Banzhaf and Deegan-Packel indexes, respectively.

Overall, it is relevant to note that no single power index represents a universal choice for measuring power in decision-making processes. As Aumann already pointed out many years ago in \cite{aumann_power_indexes}, distinct contexts may require different power indexes. Even within the same context, alternative indexes may highlight diverse aspects of the voting process, such as the probability of influencing outcomes, the decisiveness of a vote, or the potential for forming coalitions. Therefore, to make better use of power indexes, it is essential to provide an axiomatic characterization for each of them, detailing the  mathematical properties that distinguish them.

In this paper, we introduce a modification of the Felsenthal index enriched with a priori unions, which will be called the \textit{Felsenthal Owen index}. Additionally, we provide two axiomatic characterizations of this new index, combining established and novel properties for coalitional power indexes. Finally, the Felsenthal Owen index is employed to offer a detailed analysis of power distribution within the International Monetary Fund (IMF), based on its structure as of March 2025.

The paper is organized as follows. Section 2 introduces the Felsenthal Owen index, including its game-theoretic background. Section 3 details two distinct axiomatic characterizations of the proposed index. In Section 4, the Felsenthal Owen index is used to study the allocation of power within the IMF as of March 2025. Section 6 concludes with some general comments. An Online Resource Section (ORS) follows, containing the computations that underpin the IMF analysis presented in Section 4.  Finally, the Appendix proves the independence of the axioms used in each characterization.

\section{Preliminaries}
\subsection{Simple games}
Let $N=\{1,2,\ldots,n\}$ denote a finite set of \textit{players}, representing the members of a collective decision-making body, and $\mathcal{P}(N)$ the power set of $N$. In the latter, we will call \textit{coalition} to every subset of players $S\in \mathcal{P}(N)$. A \textit{simple game} is a pair $(N,v)$ where $v:\mathcal{P}(N)\rightarrow \mathbb{R}$ is a function satisfying:
\begin{itemize}[label=\scalebox{0.75}{$\bullet$}]
    \item  $v(S) \in \{0,1\}$  $ \forall S \subseteq N$
    \item $v$ is monotone: $S \subseteq T \subseteq N \Rightarrow v(S) \leq v(T)$
    \item $v(\emptyset)=0$ and $v(N)=1$.
\end{itemize}

 We denote by $SI(N)$ the set of simple games with set of players $N$.  Equivalently, a simple game can be defined as a pair $(N,W)$, where $W$ is the set of \textit{winning coalitions} $W=\{S\subseteq N \mid v(S)=1 \}$, which are precisely those coalitions capable of approving a proposal. Furthermore, due to monotonicity, $(N,W)$ can be described by the subset of \textit{minimal winning coalitions} $W^{m} =\{ S\in W \mid T \subsetneq S \Rightarrow T \notin W \}.$ We denote the subset of \textit{winning coalitions of least size} as:
 $$
W^{ls} =\{ S\in W \mid |T| < |S|  \Rightarrow T \notin W \} \subseteq W^{m}.
$$

For subsets containing a player $i\in N$ we will use $W_{i}$, $W_{i}^{m}$, and  $W_{i}^{ls} $. \medskip

In simple games, certain players exhibit distinctive roles.  A \textit{null player} in a simple game $(N,W)$ is a player $i$ such that $W_{i}^{m}= \emptyset$. A \textit{dictator player} is a player who
constitutes the sole minimal winning coalition,  $W^{m}=\{\{i\}\}$, so that the remaining players are
null. A player $i$ who belongs to every winning coalition $W_i^{m} = W^{m}$,  is called a \textit{vetoer}. Two players $i,j \in N$ are \textit{symmetric} if $ S\cup \{i\} \in W \Leftrightarrow S\cup \{j\} \in W$ for every $S \subseteq N\setminus \{i,j\}.$ \medskip
 
We now  introduce notable types of simple games and two operations to construct a new simple game. Given a non-empty coalition $S\subseteq N$, the \textit{unanimity game of} $S$ $(N,W_{S})$ is the simple game defined by $(W_{S})^{m}=\{S\}$. 
A \textit{weighted voting game} is a simple game $(N,W)$ that can be represented by $[q; w_{1},\ldots, w_{n}]$, where $q\in \mathbb{R}^{n}$ is a fixed quota, $w_{i}\in \mathbb{R}$ is the weight for a player $i\in N$, and a subset $S\subseteq N$ is a winning coalition iff $\sum_{i\in S} w_{i} \geq q $. Given two simple games with the same set of players $W$, $V \in SI(N)$, the following operations result in new simple games; the \textit{disjunction game} $(W^{\vee}V)^{m}=W^{m} \cup V^{m}$ and the \textit{conjuction game} $(W^{\wedge} V)^{m}=W^{m} \cap V^{m}$.\medskip 
 
 A \textit{power index} is a function $f$ that assigns to every simple game $(N,W) \in SI(N)$ a vector $f(N,W)\in \mathbb{R}^{n}$, such that the $i$-th component of this vector $f_{i} (N,W)$ can be interpreted as the power of
player $i$ in the game $(N,W)$ according to $f$.

\subsubsection{The Felsenthal power index}
In a multitude of collective decision-making entities, leadership roles are often established through the process of voting. This is observed in various settings such as the formation of government in parliamentary systems, the election of community leaders in neighbourhoods, or the appointment of board members in companies. Felsenthal argues in \cite{7_felsenthal_power_index} that in such contexts, the coalitions that emerge victorious from a vote are not only minimal-excluding any player whose inclusion is unnecessary for victory- but also of the smallest possible size. To illustrate this point, let us consider a four-player game $N=\{1,2,3,4\}$ with minimal winning coalitions $W^{m}=\{\{2,3,4\},\{1,2\}\}$. In this case, player 2 would never form a three-member coalition $\{2,3,4\}$ when a two-member coalition $\{1,2\}$ suffices to secure a majority. This strategic preference arises from the assumption that, once a winning coalition is formed, the total power (normalized to 1) is distributed among its members. Consequently, each member maximizes his/her individual power share by minimizing the number of participants. Moreover, Felsenthal suggests that consensus building is generally easier among fewer participants. Under this premise, the Felsenthal power index allocates power exclusively among players who participate in least sized winning coalitions. Formally, the \textit{Felsenthal power index}~\cite{7_felsenthal_power_index} of a player $i\in N $ in the simple game $(N,W)$ is given by:
$$
\psi_{i}(N,W) = \frac{1}{|W^{ls}|}\sum_{S\in W^{ls}_{i}}\frac{1}{|S|}.
$$
The power index operates under the assumption that all least size winning coalitions have an equal chance of forming $|W^{ls}|^{-1}$, and that the power within each one is evenly distributed among all members, i .e., each player in $S\in W^{ls}$ gets  $|S|^{-1}$. To simplify notation, we will occasionally denote $ p_{W}=|W^{ls}|$ relating to the \textit{probability} of each possible winning outcome and $ c_{W}=|S|$, where $S \in W^{ls}$ relating to the \textit{contribution} or share of each player in the winning coalition that finally emerges victorious.  The index can be rewritten as:
$$
\psi_{i}(N,W) = \frac{|W^{ls}_{i}|}{p_W\cdot c_{W}}.
$$

\subsection{Simple games with a priori unions}

In \cite{25_owen_value}, Owen proposed a novel model incorporating the tendency of certain players to cooperate more frequently than others. Formally, a \textit{simple game with a priori unions} is a triple $(N,W,P)$, where $(N,W)$ is a simple game and $P=\{P_{1},\ldots, P_{u}\}$ is a partition of $N$. There are two trivial a priori unions for player set $N$. The first is the structure where each player forms his own union, $N^{0}=\{\{1\},\ldots,\{n\}\}$.
The second one,  $N^{N}=\{N\}$, consists only of the grand union. We denote by $SIU(N)$ the set of simple games with a priori unions with set of players $N$.  Owen's model formalizes a two-level bargaining process.\medskip

The first level, \textit{inter-union bargaining}, captures negotiations among unions and is modelled by the \textit{quotient game}. Given a simple game with a priori unions $(N,W,P)\in SIU(N)$, the \textit{quotient game} of $(N,W,P)$ is the simple game $(U,\overline{W})$, where the players in $U$ are the unions of $P$ and
 $$\overline{W}= \left\{ R \subseteq U \mid \textstyle\bigcup\limits_{k\in R}P_{k} \in W \right\}.$$
 
For a coalition $S\subseteq N$, we define $u(S)$ as the set of representatives of $S$ in the game $(U,\overline{W})$, i.e., $u: \mathcal{P}(N) \rightarrow \mathcal{P}(U)$, $u(S)= \{k\in U \mid P_{k}\cap S \neq \emptyset \}$ for $S\subseteq N$.  We will say that a minimal winning coalition $S\in W^{m}$ is \textit{irrelevant} if its representative in the quotient game is not minimal, i.e., $u(S)\notin \overline{W}^{m}$, where $\overline{W}^{m}=\left\{ R\in \overline{W} \mid R' \subsetneq R \Rightarrow R' \notin \overline{W} \right\}.$ Given a simple game with a priori unions $(N,W,P)\in SIU(N)$, we say that a union is  \textit{null, dictator} or \textit{vetoer} in $(N,W,P)$ when it holds the corresponding role in the simple quotient game $(U,\overline{W}).$ \medskip

The second level, \textit{intra-union bargaining}, involves deliberations among members within each union. Once unions agree on their shares in the quotient game, each union $P_{k}$ internally negotiates how to distribute its gains. Owen models this as an internal game where coalitions $S\subseteq P_{k}$ bargain with their partners $P_{k} \backslash S$, considering what they can achieve by themselves and cooperation with other unions  $\cup_{l\in U\backslash \{k\}} P_{l}$, without the help of $P_{k} \backslash S$.  It is important to note that Owen does not account for the possibility of a subset $S\subseteq P_{k}$
forming alliances with \textit{proper} subsets of other unions $\widetilde{S}\subsetneq P_{l}$, $l\in U\backslash \{k\}$. Building on this approach, we define for each least size winning coalition in the quotient game $R\in \overline{W}^{ls}$, $R\neq \emptyset$, and union $k\in R$, the internal simple game $(P_{k},W_{R,k})$  conducted by players of $P_{k}$ with set of winning coalitions:
$$W_{R,k}=\left\{S\subseteq P_{k} \mid S \cup \left(\cup_{l\in R\setminus k}P_{l}
\right)\in W \right\}.$$

We designate to the subset of minimal winning coalitions in $(P_{k},W_{R,k})$ as the set of \textit{essential coalitions of}  $k$ \textit{with respect to} $R$, denoted as $E^{m}_{R,k}(N,W,P):=(W_{R,k})^{m}$. Accordingly, the subset of least size winning coalitions is referred to as the set \textit{essential coalitions of least size of} $k$ \textit{with respect to} $R$ and denoted by\footnote{Occasionally, when clear from context, we will omit the explicit expression of the simple game writing $E_{R,k}^{ls}$.}: 
\begin{equation*}
\begin{split}
E_{R,k}^{ls}(N,W,P):=(W_{R,k})^{ls} = \bigg\{ & S \subseteq P_{k} \mid S \cup \left(\cup_{l\in R\setminus \{k\}}P_{l}\right)\in W, \\
& T \cup \left(\cup_{l\in R\setminus \{k\}}P_{l}\right)\notin W\ \forall T \subset P_{k}, |T|< |S| \bigg\}.
\end{split}
\end{equation*}
For subsets containing a player $i\in N$, we will use $W_{R,k,i}$,  $E_{R,k,i}^{m}(N,W,P)$, and $E_{R,k,i}^{ls}(N,W,P)$. In addition, we define the set of essential coalitions of least size of the game $(N,W,P)$ as the union: $$E^{ls}(N,W,P)= \textstyle\bigcup\limits_{R\in \overline{W}^{ls}}\textstyle\bigcup\limits_{k\in R} E_{R,k}^{ls}(N,W,P).$$\medskip  

To illustrate the previous content, we present the following example.
\begin{example}\label{ej:esenciales1}
 Consider the simple game with a priori unions $(N,W,P)$ with  $N=\{a,b,c,d,e,f,g\}$, $P=\{P_{1},P_{2},P_{3}\}$  where $P_{1}=\{a,b,c\}$, $P_{2}=\{d,e,f\}$, $P_{3}=\{g\}$,  and $W^{m}=\{\{a,b,f\},\{a,c,f\},\{a,b,c,d\},\{a,g\},\{e,g\}\}$. Then, the set of least size winning coalitions in the quotient game is $\overline{W}^{ls}= \{\{1,2\},\{1,3\},\{2,3\}\}$ and the sets of essential coalitions of least size are:
\begin{align*}
E^{ls}_{\{1,2\},1}&=\{\{a,b\},\{a,c\}\}          &    E^{ls}_{\{1,3\},1}&=\{\{a\}\}  &  E^{ls}_{\{2,3\},2}&=\{\{e\}\} \\   E^{ls}_{\{1,2\},2} &=\{\{d\},\{f\}\}           &
E^{ls}_{\{1,3\},3}&=\{\{g\}\}  &  E^{ls}_{\{2,3\},3}&=\{\{g\}\}.
\end{align*}
\end{example}

In line with the definition of power index, a \textit{coalitional power index} is a function $F$ that assigns to each simple game with a priori unions $(N,W,P)\in SIU(N)$ an $n$-dimensional real vector $F(N,W,P)\in \mathbb{R}^{n}$. 

\subsubsection{The Felsenthal Owen power index}
The \textit{Felsenthal Owen power index}  of a player $i\in P_{k}$ in the simple game with a priori unions $(N,W,P)$ is given by
\begin{equation*} \label{eq:Indice_Owen_Felsenthal}
\Psi_{i}(N,W,P)= \frac{1}{|\overline{W}^{ls}|}\sum_{R\in \overline{W}^{ls}_{k}}\frac{1}{|R|}\frac{1}{|E_{R,k}^{ls}(N,W,P)|} \sum_{S\in E_{R,k,i}^{ls}(N,W,P)}\frac{1}{|S|} .
\end{equation*}
The index can alternatively be expressed in terms of the Felsenthal power index $\psi$:

\begin{equation*}
    \Psi_{i}(N,W,P)= \frac{1}{|\overline{W}^{ls}|}\sum_{R\in \overline{W}^{ls}_{k}}\frac{1}{|R|}\psi_{i}(P_{k},W_{R,k}).
\end{equation*}
 This index coincides with the Felsenthal index when the a priori union structure is trivial, i.e., when it corresponds to either $N^{0}$ or $N^{N}$.  Consequently, the Felsenthal Owen index generalizes the Felsenthal index incorporating the possibility of predefined coalition structures.\\
 As with the Felsenthal index, the Felsenthal Owen index is also subject to a probabilistic interpretation. The index can be understood as the expected contribution of a player within a two-stage probabilistic model, where coalition formation is restricted to equiprobable least size minimal winning coalitions at both the inter-union and intra-union levels and where power is evenly distributed between members of each least size winning coalition.\\
 \begin{example}
 Let us compute the Felsenthal Owen index for player $b\in P_{1}$ in the game $(N,W,P)$ defined in Example~\ref{ej:esenciales1}. Player $b$ belongs to union $P_{1}$,  which is part of two least size winning coalitions in the quotient game: $\{1,2\},\{1,3\}\in \overline{W}^{ls}$. However, $b$ is only relevant in the internal game associated with $\{1,2\}$, where it contributes to exactly one least size essential coalition, $\{a,b\} \in E^{ls}_{\{1,2\},1}$. The Felsenthal Owen index for $b$ is computed as:
$$
     \Psi_{b}(N,W,P)= \frac{1}{|\overline{W}^{ls}|} \frac{1}{|\{1,2\}|} \frac{1}{|E^{ls}_{\{1,2\},1}|} \frac{1}{|\{a,b\}|} =\frac{1}{3}\frac{1}{2}\frac{1}{2} \frac{1}{2}=\frac{1}{24}. $$
This calculation has a natural probabilistic interpretation. First, the probability that $\{a,b\} \in E^{ls}_{\{1,2\},1}$ is the winning coalition that ultimately emerges victorious is given by,  the probability that $\{1,2\}$ forms in the quotient game, $p_{\overline{W}}^{-1}=\frac{1}{3}$, together with the probability that $\{a,b\}$  forms within $(P_{1},W_{\{1,2\},1})$, $p_{W_{\{1,2\},1}}^{-1}=\frac{1}{2}$. Second, given its formation,  $P_{1}$ holds half the power in $ \{1,2\}$, $ c_{\overline{W}}^{-1}=\frac{1}{2}$, and within   $\{a,b\}$ player $b$  also holds half the power, $ c_{W_{\{1,2\},1}}^{-1}=\frac{1}{2}$. Combining these probabilities and power distributions gives $b$'s expected share:
     $$
     \Psi_{b}(N,W,P)= \frac{1}{p_{\overline{W}}\cdot p_{W_{\{1,2\},1}}}\hspace{0.4mm} \frac{1}{ c_{\overline{W}} \cdot  c_{W_{\{1,2\},1}}} = \frac{1}{3\cdot 2}\hspace{0.4mm}   \frac{1}{2 \cdot 2}=\frac{1}{24}. $$
Applying similar reasoning to all players yields the complete power distribution: $$\Psi(N,W,P)=\left(\frac{1}{4},\frac{1}{24},\frac{1}{24},\frac{1}{12},\frac{1}{6},\frac{1}{12},\frac{1}{3} \right).$$
 \end{example}

\section{The characterizations} 

Given the wide range of power indices currently in use,  it is crucial to determine the properties each one satisfies for two main reasons. Primarily, these properties provide a deeper comprehension of the index,  revealing how it captures different aspects of power. Secondly, they  enable meaningful comparisons between different indexes by highlighting their similarities and differences.
A particularly valuable approach is to develop an axiomatic characterization for each power index, which involves establishing a minimal set of properties that uniquely define it. In this section, we provide two such axiomatic characterizations of the Felsenthal Owen power index.

\subsection{First characterization}
The first characterization presented herein expands on the property that the Felsenthal Owen index is a generalization of the Felsenthal index. We refer to any such coalitional extension as a \textit{Coalitional Felsenthal index}.
More generally, for any power index $f$,  coalitional indices that generalize $f$ are refereed to as $\textit{Coalitional $f$ indexes}$. Prior studies have explored analogous characterizations of coalitional power indexes developing on this property. For instance, the Owen value has been characterized as a coalitional Shapley value~\cite{29_Owen_CEU} and the Symmetric Coalitional Banzhaf value as a Coalitional Banzhaf value~\cite{80_symmetric_coalitional_banzhaf_pepe}. The properties that we will use in this characterization are:\medskip

\textbf{Non-negativity (NN).} A coalitional power index $F$ satisfies NN if for every $(N,W,P) \in SIU(N)$ and $i \in N$, 
    $$F_{i}(N,W,P)\geq 0.$$

\textbf{Coalitional Felsenthal index (CFI).} A coalitional power index $F$ satisfies CFI\footnote{ Throughout this text, ``is a CFI'' and \textquotedblleft satisfies CFI property'' will be used interchangeably.} if for every $(N,W) \in SI(N)$ and $i\in N$,
     $$F_{i}(N,W,N^{0})=\psi_{i}(N,W).$$ 

\textbf{Quotient game (QG).} A coalitional power index $F$ satisfies QG if for every $(N,W,P) \in SIU(N)$ and $k\in U$,
     $$F_{k}(U,\overline{W},U^0)=\sum_{i\in P_{k}}F_{i}(N,W,P).$$ 

\textbf{Proportionality with respect to essential coalitions of least size (PELS).}  A coalitional power index $F$ satisfies PELS if, for every $(N,W,P)\in SIU(N)$ and every $i,j\in P_{k}\in P$, 
    \begin{equation*}\label{eq:property_prop_essential_coalitions}
          F_{i}(N,W,P)\sum\limits_{R\in \overline{W}^{ls}_{k}}  F_{j}(P_{k},W_{R,k},P_{k}^{0})=  F_{j}(N,W,P)\sum\limits_{R\in \overline{W}^{ls}_{k}}  F_{i}(P_{k},W_{R,k},P_{k}^{0}).
    \end{equation*}
Loosely speaking, this property establishes that, for players $i$ and $j$ belonging to the same union $P_{k}$, the relative proportion of power between them in the global game $(N,W,P)$ is equal to the aggregate of their power proportions across all internal games $(P_{k},W_{R,k},P_{k}^{0})$ for $R\in \overline{W}^{ls}$. In essence, this property reflects that the power-based ranking of players of the same union $P_{k}$ in the global game $(N,W,P)$, is entirely determined by their performance in the internal subgames. Moreover, their power in the global game is a re-scaled version of what they obtain in the internal subgames, with the same scaling factor applied to all union members.



\begin{theo}
\label{Theorem1}
The Felsenthal Owen power index is the only CFI that satisfies NN, QG, and PELS. 
\end{theo}

\begin{proof}
\quad\\
\noindent\textbf{Existence.} 
It is immediate from its definition that the Felsenthal Owen power index satisfies NN.\medskip

To show that it's a CFI power index, let us consider $(N,W,N^{0})$ and its quotient game $(U,\overline{W})$. 

 Since $N^{0}$ is the trivial union, it follows that $|\overline{W}^{ls}|=|W^{ls}|$. Let us fix $i 
 \in N$ and $k\in U$ such that $i\in P_k$. It means that $P_k=\{i\}$ and there is a bijection between $\overline{W}^{ls}_{k}$ and $W^{ls}_{i}$, such that 
 assigns to each $R\in \overline{W}^{ls}_{k}$ the coalition $S=\{i\in P_k \mid k\in R\}\in W^{ls}_{i}$, which satisfies that $|S|=|R|$. Furthermore, given $R\in \overline{W}^{ls}$, the game $(P_k,W_{R,k})$ has just one player and the Felsenthal index is efficient so $\psi_{i}(P_k,W_{R,k})=1$. Then,
    $$
    \Psi_{i}(N,W,N^{0})=\frac{1}{|\overline{W}^{ls}|}\sum_{R\in \overline{W}^{ls}_{k}}\frac{1}{|R|}\psi_{i}(P_k,W_{R,k})=\frac{1}{|W^{ls}|}\sum_{S \in W^{ls}_{i}}\frac{1}{|S|}=\psi_{i}(N,W).
    $$   

The QG property follows from the definition of $\Psi$ and the efficiency of the Felsenthal index. Indeed, for a fixed $P_k\in P$, it holds that,

\begin{align*}
    \sum_{i\in P_{k}}\Psi_{i}(N,W,P) & =\sum_{i\in P_{k}}\frac{1}{|\overline{W}^{ls}|}\sum_{R\in \overline{W}^{ls}_{k}}\frac{1}{|R|}\psi_{i}(P_{k},W_{R,k})&\\
         & =\frac{1}{|\overline{W}^{ls}|}\sum_{R\in \overline{W}^{ls}_{k}}\frac{1}{|R|}\sum_{i\in P_{k}}\psi_{i}(P_{k},W_{R,k})&\\
         & =\frac{1}{|\overline{W}^{ls}|}\sum_{R\in \overline{W}^{ls}_{k}}\frac{1}{|R|} = \Psi_{k}(U,\overline{W},U^{0}).
    \end{align*}
Lastly, we show that $\Psi$ satisfies PELS. First, note that for a player $i\in P_{k}$, $\Psi_{i}(N,W,P)\neq 0 $ if and only if there exists a coalition $R\in \overline{W}^{ls}_{k}$ such that $E^{ls}_{R,k,i}(N,W,P)\neq \emptyset$, which is equivalent to $\psi_{i}(P_{k},W_{R,k})\neq 0$. Furthermore, by the CFI property:
\begin{equation*}
    \Psi_{i}(N,W,P)= \frac{1}{|\overline{W}^{ls}|}\frac{1}{|R|}\sum_{R\in \overline{W}^{ls}_{k}}\Psi_{i}(P_{k},W_{R,k},P_{k}^{0}), 
\end{equation*}
Hence,  
$\Psi_{i}(N,W,P) \neq 0$ if and only if $\Psi_{i}(P_{k},W_{R,k},P_{k}^{0}) \neq 0$ for some $R\in\overline{W}^{ls}_{k} $.\\
Now consider two players $i,j\in P_{k}$. If  $|P_{k}|=1$, the PELS condition holds trivially. Otherwise, suppose $i,j$ are distinct. If either $\Psi_{i}(N,W,P)$ or $ \Psi_{j}(N,W,P)$ is 0 in $(N,W,P)$, the condition is again satisfied since both sides of the equality are zero as just noted. Otherwise, in the non-trivial case:
\begin{equation*}
    \frac{\Psi_{i}(N,W,P)}{\sum_{R\in \overline{W}^{ls}_{k}}\Psi_{i}(P_{k},W_{R,k},P_{k}^{0})}= \frac{1}{|\overline{W}^{ls}|}\frac{1}{|R|}=\frac{\Psi_{j}(N,W,P)}{\sum_{R\in \overline{W}^{ls}_{k}}\Psi_{j}(P_{k},W_{R,k},P_{k}^{0})}.
\end{equation*}
Thus, the PELS property is satisfied by $\Psi$.

\noindent\textbf{Uniqueness.}
We demonstrate that the Felsenthal Owen index $\Psi$ is the only coalitional index satisfying the above properties. To this aim, let $F$ be a coalitional power index  satisfying those properties and consider $(N,W,P)\in SIU(N)$ with the corresponding quotient game $(U,\overline{W})$. \medskip

\noindent Since $F$ is a CFI that satisfies QG, for each union $k\in U$:
\begin{equation}
\label{eq:charac1_unic_1}
\sum\limits_{i\in P_{k}} F_{i}(N,W,P)=F_{k}(U,\overline{W},U^{0})
=\psi_{k}(U,\overline{W})=\frac{1}{|\overline{W}^{ls}|}\sum\limits_{R\in \overline{W}^{ls}_{k}}\frac{1}{|R|}.
 \end{equation}

\noindent If $ \overline{W}^{ls}_{k}= \emptyset$, then equation~(\ref{eq:charac1_unic_1}) implies $\sum\limits_{i\in P_{k}} F_{i}(N,W,P)=0$ and, by NN it follows that $F_{i}(N,W,P)=0$ $\forall i \in P_{k}$.\medskip

\noindent Otherwise, given $R\in \overline{W}^{ls}_{k}$ and $i\in P_{k}$, CFI yields:
       \begin{equation}\label{eq:charac1_unic_2}
        F_{i}(P_{k},W_{R,k},P^{0}_{k})=\psi_i(P_{k},W_{R,k})=\frac{1}{|E^{ls}_{R,k}(N,W,P)|}\sum\limits_{S\in E^{ls}_{R,k,i}(N,W,P)}\frac{1}{|S|}.
       \end{equation} 

\noindent Now, players  in $P_{k}$ can be classified into two subsets:  $P_{k}=P_{k}^{1}\sqcup P_{k}^{2}$, where  $P_{k}^{1}$ consists of players $i\in P_{k}$ such that $E^{ls}_{R,k,i}(N,W,P)=\emptyset$ for all  $R \in \overline{W}^{ls}_{k}$ and $P_{k}^{2}=P_{k}\setminus P_{k}^{1}$. Note that $P_{k}^{2}\neq \emptyset$ as  $\overline{W}^{ls}_{k}\neq \emptyset$. This partition implies that every player  $i\in P_k$ belongs to exactly one of the two subsets:
\begin{enumerate}
\item $i\in P_{k}^{1}$. According to~(\ref{eq:charac1_unic_2}), $F_{i}(P_{k},W_{R,k},P_{k}^{0})=0$ for every  $R \in \overline{W}^{ls}_{k}$. Choosing any $j\in P_{k}^{2}$, the PELS property implies:
       $$0=F_{j}(N,W,P)\cdot 0=  F_{i}(N,W,P)\sum\limits_{R\in \overline{W}^{ls}_{k}}  F_{j}(P_{k},W_{R,k},P_{k}^{0}).
    $$
\noindent Since $j\in P_{k}^{2}$, by~(\ref{eq:charac1_unic_2}) there exists $\widetilde{R}\in \overline{W}^{ls}_{k}$ such that $F_{j}(P_{k},W_{\widetilde{R},k},P_{k}^{0})> 0$, leading to $ F_{i}(N,W,P)=0$.

\item $i\in P_{k}^{2}$. \begin{itemize}[label=\scalebox{0.75}{$\bullet$}]
    \item If $|P_{k}^2|=1$, then $P_{k}^{2}=\{i\}$ and:
$$
F_{i}(N,W,P)=\sum\limits_{j\in P_{k}}F_{j}(N,W,P)\overset{(\ref{eq:charac1_unic_1})}{=}\frac{1}{|\overline{W}^{ls}|}\sum\limits_{R\in \overline{W}^{ls}_{k}}\frac{1}{|R|}.
$$
\item If  $|P_{k}^{2}|>1$, then for any player $j\in P_{k}^{2}$, the  PELS property implies that there exist a constant $\lambda_{k}$ such that:
\begin{equation}\label{eq:charac1_unic_3}
   F_{j}(N,W,P) = \lambda_{k}   \sum\limits_{R\in \overline{W}^{ls}_{k}}F_{j}(P_{k},W_{R,k},P_{k}^{0}).
\end{equation}
Subsequently:
\begin{align*}
      \frac{1}{|\overline{W}^{ls}|}\sum\limits_{R\in \overline{W}^{ls}_{k}}\frac{1}{|R|}&\overset{(\ref{eq:charac1_unic_1})}{=}\sum\limits_{j\in P_{k}} F_{j}(N,W,P)=\sum\limits_{j\in P_{k}^{2}} F_{j}(N,W,P)&\\
      &\overset{(\ref{eq:charac1_unic_2}),(\ref{eq:charac1_unic_3})}{=} \lambda_{k}\sum\limits_{R\in \overline{W}^{ls}_{k}} \frac{1}{|R|}  \frac{1}{|E^{ls}_{R,k}(N,W,P)|}\sum\limits_{j\in P_{k}} \sum\limits_{S\in E^{ls}_{R,k,j}(N,W,P)}\frac{1}{|S|}&\\
      & =\lambda_{k}\hspace{1mm}|\overline{W}^{ls}_{k}|.
\end{align*}
Hence, $\lambda_{k}= \frac{1}{|\overline{W}^{ls}| \hspace{0.5mm}|R|}$, $R \in \overline{W}^{ls}$, and equation~(\ref{eq:charac1_unic_3}) implies that:
\begin{equation}
   F_{i}(N,W,P) = \frac{1}{|\overline{W}^{ls}|} \sum\limits_{R\in \overline{W}^{ls}_{k}} \frac{1}{|R|} F_{i}(P_{k},W_{R,k},P_{k}^{0}).
\end{equation}
Lastly, the index $ F_{i}(P_{k},W_{R,k},P_{k}^{0})$ can be substitute using equation~(\ref{eq:charac1_unic_2}).
\end{itemize}
\end{enumerate}
Thus, in all scenarios, the coalitional index $F$ coincides with the Felsenthal Owen index $\Psi$, proving it is the unique coalitional index satisfying the specified properties.

\end{proof}

\subsection{Second characterization}
 Our second characterization builds directly upon the framework developed by Freixas and Samaniego in \cite{3_Felsenthal_Freixas_samaniego}. In their work, these authors demonstrate that the Felsenthal index is the only power index satisfying the next four properties:\medskip

\textbf{Efficiency (E).} A  power index $f$ satisfies E if for every $(N,W)\in SI(N)$, $\sum_{i\in N} f_{i}(N,W) =1$.\medskip

\textbf{Null player (NP). }
        A  power index $f$ satisfies NP if for every $(N,W) \in SI(N)$ and $i\in N$ such that is a null player in the game $(N,W)$, then $f_{i}(N,W)=0$.
\medskip

\textbf{Symmetry (S). }
        A  power index $f$ satisfies S if for every $(N,W) \in SI(N)$ and $i,j\in N$ such that they are symmetric players in the game $(N,W)$, then $f_{i}(N,W)=f_{j}(N,W)$.
\medskip

\textbf{Transfer for coalitions of least size (TCLS). }
        A power index $f$ satisfies TCLS if for any pair $(N,W)$ and $(N, V)\in SI(N)$ such that $W^{ls}\cap V^{ls} =\emptyset$,  then
    $$
    f (N,W \vee V)= \left\{ \begin{array}{lcc} f (N,W), & \textrm{if} &  c_{\hspace{0.4mm}W} <  c_{\hspace{0.4mm}V} \\ \\ f (N,V), & \textrm{if} & c_{\hspace{0.4mm}V} <  c_{\hspace{0.4mm}W}\\ \\ \frac{ p_{\hspace{0.4mm}W}}{ p_{\hspace{0.4mm}W}+ p_{\hspace{0.4mm}V}}f (N,W) + \frac{ p_{\hspace{0.4mm}V}}{ p_{\hspace{0.4mm}W}+ p_{\hspace{0.4mm}V}} f(N,V),&  & \textrm{otherwise.} \end{array} \right.
    $$
This last property TCLS is closely related to the ``Transfer axiom'' originally introduced in the Shapley-Shubik characterization by Dubey~\cite{74_caracterizacion_Shapley_shubik}, although it focuses exclusively on least size winning coalitions as emphasized  by Felsenthal. In the specific case where $W$ and $V$ have disjoint minimal winning coalitions $W^{ls}\cap V^{ls} =\emptyset$ and identical least sizes $c_{\hspace{0.4mm}W} =  c_{\hspace{0.4mm}V}$, the TCLS property states that power in the disjunction game $W \vee V$ is given by a weighted average of the powers in the individual games. The weights correspond to each game’s relative contribution to the combined set of minimal winning coalitions $W^{ls}\sqcup V^{ls}$.\\[3pt]

 
 
\medskip
Next we list out several properties for coalitional power indexes. Many of them represent adaptations of the previously discussed properties (E, NP, S, and TCLS). As we will subsequently demonstrate, these adapted properties serve to uniquely characterize the Felsenthal Owen power index.\medskip

\textbf{Efficiency (E).} A coalitional power index $F$ satisfies E if for every $(N,W,P)\in SIU(N)$, $\sum_{i\in N} F_{i}(N,W,P) =1$.\medskip

\textbf{Null player (NP). }
        A coalitional power index $F$ satisfies NP if for every $(N,W,P) \in SIU(N)$ and $i\in N$ such that is a null player in the game $(N,W)$, then $F_{i}(N,W,P)=0$.

\medskip

\textbf{Symmetry among unions (S-AU).} A coalitional power index $F$ satisfies S-AU if for every $(N,W,P) \in SIU(N)$ and $k,l\in U$ 
 such that they are symmetric players in the quotient game $(U,\overline{W})$, then
     $$
    \sum_{i\in P_{k}} F_{i}(N,W,P)=\sum_{i\in P_{l}} F_{i}(N,W,P).
     $$

\textbf{Symmetry inside unions (S-IU).} A coalitional power index $F$ satisfies S-IU if for every $(N,W,P) \in SIU(N)$ and $i,j\in P_{k}\in P$, such that they are symmetric players in the game $(N,W)$, then $F_{i}(N,W,P)=F_{j}(N,W,P)$.\medskip

\textbf{Transfer for coalitions of least size among unions (TCLS-AU).} A coalitional power index $F$ satisfies TCLS-AU if for any pair $(N,W, P)$ and $(N, V, P)\in SIU(N)$ such that $\overline{W}^{ls}\cap \overline{V}^{ls} =\emptyset$,  then
    $$
    F (N,W \vee V, P)= \left\{ \begin{array}{lcc} F (N,W, P), & \textrm{if} &  c_{\hspace{0.4mm}\overline{W}} <  c_{\hspace{0.4mm}\overline{V}} \\ \\ F (N,V, P), & \textrm{if} & c_{\hspace{0.4mm}\overline{V}} <  c_{\hspace{0.4mm}\overline{W}}\\ \\ \frac{ p_{\hspace{0.4mm}\overline{W}}}{ p_{\hspace{0.4mm}\overline{W}}+ p_{\hspace{0.4mm}\overline{V}}}F (N,W, P) + \frac{ p_{\hspace{0.4mm}\overline{V}}}{ p_{\hspace{0.4mm}\overline{W}}+ p_{\hspace{0.4mm}\overline{V}}} F (N,V, P),&  & \textrm{otherwise.} \end{array} \right.
    $$

\textbf{Transfer for coalitions of least size inside unions (TCLS-IU).} A coalitional power index $F$ satisfies TCLS-IU if for any pair $(N,W, P)$ and $(N, V, P)\in SIU(N)$ verifying that  $W^{ls}\cap V^{ls} =\emptyset$ and there exists $k\in U$ such that for every $S\in W^{m} \cup V^{m}$ it holds that $S\subseteq P_{k}$, then
       $$
    F(N,W \vee V, P)= \left\{ \begin{array}{lcc} F (N,W, P), & \textrm{if} &  c_{\hspace{0.4mm}W} <  c_{\hspace{0.4mm}V} \\ \\ F (N,V, P), & \textrm{if} & c_{\hspace{0.4mm}V} <  c_{\hspace{0.4mm}W}\\ \\ \frac{ p_{\hspace{0.4mm}W}}{ p_{\hspace{0.4mm}W}+ p_{\hspace{0.4mm}V}}F (N,W, P) + \frac{ p_{\hspace{0.4mm}V}}{ p_{\hspace{0.4mm}W}+ p_{\hspace{0.4mm}V}} F (N,V, P),&  & \textrm{otherwise.} \end{array} \right.
    $$

\textbf{Independence of irrelevant coalitions (IIC).} A coalitional power index $F$ satisfies IIC if for any $(N,W,P)$, given $\widetilde{S} \in W^{m}$ such that $u(\widetilde{S}) \notin \overline{W}^{m}$, then $F(N,W,P)= F(N,W',P)$ where $(W')^m = W ^m \setminus \{\widetilde{S}\}$.\\
This property says that  only coalitions whose representatives play an essential role in the quotient game affect power distribution.\medskip

\textbf{Invariance with respect to least size essential coalitions (ILSE).} A coalitional power index $F$ satisfies ILSE if for any pair $(N,W, P)$ and $(N, V, P) \in SIU(N)$, verifying that $ E^{ls}(N,W, P) = E^{ls}(N, V, P)$ and that there exists $R=\{k_{1},\ldots,k_{s}\}\subseteq U$ such that $\overline{W}^{ls} = \{R\}$ and $\overline{V}^{ls} = \{\{k_{1}\},\ldots,\{k_{s}\}\}$,  then $F(N,W, P) = F (N, V, P)$. 
The ILSE property states that if a game has a unique least size winning coalition of unions, say $\overline{W}^{ls}=\{\{k_{1},\ldots,k_{s}\}\}$, the power distribution among players remains the same when the winning coalitions $S\in W^{m}$ whose representatives are $\{k_{1},\ldots,k_{s}\}$, are decomposed into individual unions
 $S\cap P_{1}, \ldots S \cap P_{s}\in V^{m}$, since essential coalitions of least size 
remain the same. This property is closely related to Owen’s approach in the two-level bargaining model, which assumes that subsets of a union cannot ally with proper subsets of other unions. We illustrate this property with the following example.\\

\begin{example}\label{ej:esenciales2}
 Consider the player set $N=\{a,b,c,d,e,f\}$, partitioned into two groups $P_{1}=\{a,b,c,d\}$ and $P_{2}=\{e,f\}$. Examine the games $(N,W_{1},P)$ and $(N,V,P)$, with the following sets of minimal winning coalitions:
$$
W_{1}^{m}=\{\{a,b,e\},\{c,d,f\},\{a,b,f\}\}  \text{ and } V^{m}=\{\{a,b\},\{c,d\},\{e\},\{f\}\}.
$$ 
Clearly,  minimal winning coalitions of $V$ are those of $W_{1}$ decomposed into individual unions. These two games are under the conditions of the ILSE property, because their quotient games verify that  $\overline{W_{1}}^{ls}=\{\{1,2\}\}$ and $\overline{V}^{ls} =\{\{1\},\{2\}\}$, and their least size essential coalitions coincide:
 $$ 
 E^{ls}(N,W_{1},P)=\{\{a,b\},\{c,d\},\{e\},\{f\}\}= E^{ls}(N,V,P).
 $$ 
Therefore, for any index $F$ satisfying ILSE property, it follows that $F(N,W_{1}, P) = F (N, V, P)$. Now, consider a third game $W_{2}$, where $W^{m}_{2}$ is formed by removing coalition $\{a,b,f\}$ from $W^{m}_{1}$: $$W^{m}_{2}=\{\{a,b,e\},\{c,d,f\}\}.$$
This game still meets the same ILSE conditions, hence $F(N,W_{1}, P) = F (N, W_{2}, P)$.

The interpretation behind this result is that coalition $\{a,b,f\}$ in $W_{1}$ does not supply any additional power to players $a$ and $b$, provided that $\{a,b,e\}$ already exists.
Even though $\{a,b\}$ appears to have two distinct ways to win in $W_{1}$ (with $e$ or with $f$) and only one in $W_{2}$, the coalition structure dictates that $a$ and $b$ can only negotiate their power by joining the entire union $P_{2}=\{e,f\}$. Hence, the availability of partial cooperation does not alter the resulting power distribution.
\end{example}


\begin{theo}
\label{Theorem2}
The Felsenthal Owen power index is the only coalitional power index that satisfies E, NP, S-AU, S-IU, TCLS-AU, TCLS-IU, IIC, and ILSE. 
\end{theo}

\begin{proof}
\quad\\
\noindent\textbf{Existence.}
First, we will briefly show that the Felsenthal Owen power index satisfies E, NP, S-AU, and S-IU.\medskip

\noindent \textit{E:} Since the Felsenthal power index is efficient and the Felsenthal Owen power index satisfies the QG property, 

$$\sum_{i\in N}\Psi_{i}(N,W,P) = \sum_{k\in U} \sum_{i\in P_{k}}\Psi_{i}(N,W,P) = \sum_{k\in U} \psi_{k}(U,\overline{W}) =1.
    $$ 

\noindent \textit{NP:} Let $i\in N$ be a null player, i.e., $W^{m}_{i}=\emptyset$. Then, for every $R\in \overline{W}^{ls}$, $E_{R,k,i}(N,W,P)=\emptyset$, what implies that $\Psi_{i}(N,W,P)=0$.\medskip 

\noindent \textit{S-AU:} Given two symmetric players $k$ and $l$ in the game $(U,\overline{W})$, the QG property and the symmetry in the case of the Felsenthal power index imply that
     $$
    \sum_{i\in P_{k}} \Psi_{i}(N,W,P)=\psi_{k}(U,\overline{W})=\psi_{l}(U,\overline{W})=\sum_{i\in P_{l}} \Psi_{i}(N,W,P).
     $$

\noindent \textit{S-IU:} Let $i,j$ be two symmetric players in the union $ P_{k}$. For each $R\in \overline{W}^{ls}_{k}$, they are symmetric players in the game $(P_{k},W_{R,k})$. Since $\psi$ satisfies the property of symmetry,    
\begin{equation*} 
\begin{split}
\Psi_{i}(N,W,P)& = \frac{1}{|\overline{W}^{ls}|}\sum_{R\in \overline{W}^{ls}_{k}}\frac{1}{|R|}\psi_{i}(P_{k},W_{R,k})  \\
 & = \frac{1}{|\overline{W}^{ls}|}\sum_{R\in \overline{W}^{ls}_{k}}\frac{1}{|R|}\psi_{j}(P_{k},W_{R,k})=\Psi_{j}(N,W,P).
\end{split}
\end{equation*}

\noindent Next, we will prove that $\Psi$ satisfies both TCLS-AU and TCLS-IU.\medskip 

\noindent\textit{TCLS-AU:} Let us consider $(N,W, P)$ and  $(N, V, P) \in SIU(N)$ such that $\overline{W}^{ls}\cap \overline{V}^{ls} =\emptyset$. There are three options:
    \begin{enumerate}[label=(\roman*)]
    \item  $c_{\hspace{0.4mm}\overline{W}} < c_{\hspace{0.4mm}\overline{V}}$. Then, $(\overline{W} \vee \overline{V})^{ls} = \overline{W}^{ls} $ and, given $R \in \overline{W}^{ls}$ and $k\in R$, $E^{ls}_{R,k}(N,W\vee V,P) = E^{ls}_{R,k}(N,W,P)$. Hence, for any $i\in P_{k}$, 
        \begin{align*}
                \mathclap{\hspace{20mm}\Psi_{i}(N,W \vee V,P)} & \notag \\
                &= \frac{1}{|\overline{W \vee V}\hspace{0.2mm}^{ls}|}\sum_{R\in \overline{W\vee V}\hspace{0.01mm}^{ls}_{k}}\frac{1}{|R|}\frac{1}{|E_{R,k}^{ls}(N,W \vee V,P)|} \sum_{S\in E_{R,k,i}^{ls}(N,W \vee V,P)}\frac{1}{|S|} \\
                &=\frac{1}{|\overline{W}\hspace{0.2mm}^{ls}|}\sum_{R\in \overline{W}\hspace{0.01mm}^{ls}_{k}}\frac{1}{|R|}\frac{1}{|E_{R,k}^{ls}(N,W ,P)|} \sum_{S\in E_{R,k,i}^{ls}(N,W ,P)}\frac{1}{|S|} \\
                &=\Psi_{i}(N,W ,P).
        \end{align*}
    \item  $c_{\hspace{0.4mm}\overline{W}} > c_{\hspace{0.4mm}\overline{V}}$. It is analogous to the previous case.
     \item  $c_{\hspace{0.4mm}\overline{W}} = c_{\hspace{0.4mm}\overline{V}}$. Then, since  $\overline{W}^{ls}\cap \overline{V}^{ls} =\emptyset$, we distinguish two possibilities for each $R\in (\overline{W} \vee \overline{V})^{ls}:
        $\begin{itemize}
            \item $R \in \overline{W}^{ls}$. For each $k\in R$, $E^{ls}_{R,k}(N,W\vee V,P) = E^{ls}_{R,k}(N,W,P)$.
            \item $R \in \overline{V}^{ls}$. For each $k\in R$, $E^{ls}_{R,k}(N,W\vee V,P) = E^{ls}_{R,k}(N,V,P)$.
        \end{itemize}
    Thus,\\[7pt]
\noindent
\(\Psi_{i}(N,W \vee V,P)\)
\vspace{-0.5em}
\begin{align}
    &= \frac{1}{|\overline{W \vee V}\hspace{0.2mm}^{ls}|}\sum_{R\in \overline{W\vee V}\hspace{0.01mm}^{ls}_{k}}\frac{1}{|R|}\frac{1}{|E_{R,k}^{ls}(N,W \vee V,P)|} \sum_{S\in E_{R,k,i}^{ls}(N,W \vee V,P)}\frac{1}{|S|} \notag \\
    &= \frac{1}{|\overline{W \vee V}\hspace{0.2mm}^{ls}|}\Bigg[\sum_{R\in \overline{W}\hspace{0.01mm}^{ls}_{k}}\frac{1}{|R|}\frac{1}{|E_{R,k}^{ls}(N,W \vee V,P)|} \sum_{S\in E_{R,k,i}^{ls}(N,W \vee V,P)}\frac{1}{|S|} \notag \\
    & \quad + \sum_{R\in \overline{V}\hspace{0.01mm}^{ls}_{k}}\frac{1}{|R|}\frac{1}{|E_{R,k}^{ls}(N,W \vee V,P)|} \sum_{S\in E_{R,k,i}^{ls}(N,W \vee V,P)}\frac{1}{|S|}\Bigg] \\
    &= \frac{1}{|\overline{W \vee V}\hspace{0.2mm}^{ls}|}\Bigg[\sum_{R\in \overline{W}\hspace{0.01mm}^{ls}_{k}}\frac{1}{|R|}\frac{1}{|E_{R,k}^{ls}(N,W ,P)|} \sum_{S\in E_{R,k,i}^{ls}(N,W ,P)}\frac{1}{|S|} \notag \\
    & \quad + \sum_{R\in \overline{V}\hspace{0.01mm}^{ls}_{k}}\frac{1}{|R|}\frac{1}{|E_{R,k}^{ls}(N, V,P)|} \sum_{S\in E_{R,k,i}^{ls}(N, V,P)}\frac{1}{|S|}\Bigg] \\
    &= \frac{ p_{\hspace{0.4mm}\overline{W}}}{ p_{\hspace{0.4mm}\overline{W}}+ p_{\hspace{0.4mm}\overline{V}}}\Psi_{i} (N,W, P) + \frac{ p_{\hspace{0.4mm}\overline{V}}}{ p_{\hspace{0.4mm}\overline{W}}+ p_{\hspace{0.4mm}\overline{V}}}\Psi_{i}(N,V, P). \notag
\end{align}

    
    \end{enumerate}
 \noindent\textit{TCLS-IU:} Given $(N,W, P)$ and $(N, V, P) \in SIU(N)$ verifying that, for every $S\in W^{m} \cup V^{m}$, it holds that there exists $P_k\in P$ such that $S\subseteq P_{k}$.
    Then $\overline{W}^{ls}=\overline{V}^{ls}=\overline{W\vee V}^{ls}= \{\{k\}\}$ and $E_{\{k\},k}^{ls}(N,W\vee V,P) = (W\vee V)^{ls}$. So
     
    $$
    \Psi_{i}(N,W \vee V,P) =\frac{1}{| (W\vee V)^{ls}|}\sum_{S\in (W\vee V)_{i}^{ls}}\frac{1}{|S|}.
    $$
    Hence, there are three options:
    \begin{enumerate}[label=(\roman*)]
        \item $c_{W}<c_{V}$. In this case, $(W\vee V)^{ls}=W^{ls}$ and
         $$
        \Psi_{i}(N,W \vee V,P)= \frac{1}{| W^{ls}|}\sum_{S\in W_{i}^{ls}}\frac{1}{|S|}=\Psi_{i}(N,W,P).
        $$
      
        \item $c_{V}<c_{W}$. Analogous to the previous case.
        \item $c_{W}=c_{V}$. As  $W^{ls}\cap V^{ls} =\emptyset$, 
         \begin{align*}
          \Psi_{i}(N,W \vee V,P) &=\frac{1}{| W^{ls}|+| V^{ls}|}\Bigg[\sum_{S\in W_{i}^{ls}}\frac{1}{|S|}  + \sum_{S\in V_{i}^{ls}}\frac{1}{|S|} \Bigg] &\\
          & =\frac{ p_{\hspace{0.4mm}W}}{ p_{\hspace{0.4mm}W}+ p_{\hspace{0.4mm}V}}\Psi_{i} (N,W, P) + \frac{ p_{\hspace{0.4mm}V}}{ p_{\hspace{0.4mm}W}+ p_{\hspace{0.4mm}V}} \Psi_{i}(N,V, P),
        \end{align*}
        where the last equality follows from the fact that $E_{\{k\},k}^{ls}(N,W,P) = W^{ls}$ and  $E_{\{k\},k}^{ls}(N,V,P) = V^{ls}$.
    \end{enumerate}
Finally, we will highlight the main steps to prove that $\Psi$ satisfies IIC and ILSE.\medskip

\noindent\textit{IIC:} If $(N,W,P)$, $(N,W',P)\in SIU(N)$ are such that $u(\widetilde{S}) \notin \overline{W}^{m}$ and $(W')^m = W ^m \setminus \{\widetilde{S}\}$, then $\overline{W}^{ls}=(\overline{W'})^{ls}$ and, for $R \in \overline{W}^{ls}=(\overline{W'})^{ls}$ and $k\in R$, $E^{ls}_{R,k}(N,W,P)=E^{ls}_{R,k}(N,W',P)$. Then
\begin{equation*} 
\begin{split}
\Psi_{i}(N,W,P)& =  \frac{1}{|\overline{W}^{ls}|}\sum_{R\in \overline{W}^{ls}_{k}}\frac{1}{|R|}\frac{1}{|E_{R,k}^{ls}(N,W,P)|} \sum_{S\in E_{R,k,i}^{ls}(N,W,P)}\frac{1}{|S|}   \\
 & = \frac{1}{|(\overline{W'})^{ls}|}\sum_{R\in (\overline{W'})^{ls}_{k}}\frac{1}{|R|}\frac{1}{|E_{R,k}^{ls}(N,W',P)|} \sum_{S\in E_{R,k,i}^{ls}(N,W',P)}\frac{1}{|S|} \\
 & = \Psi_{i}(N,W',P).
\end{split}
\end{equation*}

\noindent\textit{ILSE:} In this case, for each $k_{j}\in R$, with $j\in \{1,\ldots,s\}$, we know that $E^{ls}_{R,k_{j}}(N,W,P)=E^{ls}_{\{k_{j}\},k_{j}}(N,V,P)$. Since the Felsenthal index $\psi$ matches for games with the same set of least size winning subsets, we deduce that, for $i \in P_{k_{j}}$,  $\psi_{i}(P_{k_{j}},W_{R,k_{j}})=\psi_{i}(P_{k_{j}},V_{\{k_{j}\},k_{j}})$. So,
$$
\Psi_{i}(N,W,P) = \frac{1}{|R|}\psi_{i}(P_{k},W_{R,k_{j}})= \frac{1}{|\overline{V}^{ls}|}\psi_{i}(P_{k},V_{\{k_{j}\},k_{j}})=\Psi_{i}(N,V,P) .
$$

\noindent\textbf{Uniqueness: }  We demonstrate that the Felsenthal Owen index $\Psi$ is the only coalitional index satisfying the above properties. To this aim, let $F$ be a coalitional power index satisfying those properties and consider $(N,W,P)\in SIU(N)$ with set of minimal winning coalitions $W^{m}=\{S_{1},\ldots,S_{q}\}$. Since $F$ satisfies IIC, we can assume that 
$u(S) \in \overline{W}^{m}$, for every $ S\in W^{m}$.

\noindent We begin by considering the case when $q=1$. In this scenario, $W^{m} = \{S\}=W^{ls}$ for a certain set $S\subseteq N$, i.e., $(N,W)$ is the unanimity game of $S$. Consequently,  the quotient game $(U,\overline{W})$ is the unanimity game of $u(S)$ since $\overline{W}^{m}=\{u(S)\}=\overline{W}^{ls} $. By E and NP:
\begin{equation*}
   1=\sum_{i\in N}F_{i}(N,W,P) = \sum_{i\in S}F_{i}(N,W,P)=\sum_{k\in u(S)}\sum_{i\in S\cap P_{k}}F_{i}(N,W,P).
\end{equation*}

 \noindent Moreover, since the unions in $u(S)$ are symmetric in $(U,\overline{W})$ and the players in $S\cap P_{k}$ are  symmetric in $(N,W)$, by S-AU and S-IU we can conclude that for each $i\in P_k\in P$:
\begin{equation}\label{eq:charac2_unic_1}
  F_{i}(N,W,P) = \begin{cases}
         \frac{1}{|u(S)|| S\cap P_{k}|} & ,\text{ if} \ i \in S\cap P_{k}\\
        0 & , \text{ if} \ i \notin S\cap P_{k}.
    \end{cases}
\end{equation}

\noindent Now, let us examine the case where $q>1$. To this aim, consider a partition of $W^{m}$ into sets $\{T_{1},\ldots, T_{l}\}$, defined by the equivalence relation:
$$S_{d_{1}}, S_{d_{2}} \in T_{h} \Leftrightarrow u(S_{d_{1}})= u( S_{d_{2}}) = R_{h},\hspace{2mm}h \in \{1,\ldots,l\}.$$ 
\noindent Thus, each  $T_{h}$ consist of coalitions $ T_{h} = \{ S_{h_{1}},S_{h_{2}},\ldots, 
S_{h_{t_{h}}}\}\subseteq W^{m}$, and we decompose $W$ as follows:
$$W = W^{T_{1}} \vee \ldots \vee  W^{T_{l}},\text{ where }    W^{T_{h}}=  W_{S_{h_{1}}} \vee \ldots \vee  W_{S_{h_{t_{h}}}}.
$$\medskip

\noindent Given that $\overline{W}^{m} =\{R_{1},\ldots, R_{l}\}$, and assuming $\overline{W}^{ls} =\{R_{1},\ldots, R_{l'}\}$, with $l'\le l$. Then, we can rewrite:
  \[W=\underbrace{W^{T_{1}} \vee \ldots \vee  W^{T_{l'}}}_{V_{1}}\vee\underbrace{W^{T_{l'+1}} \vee \ldots \vee  W^{T_{l}}}_{V_{2}} = V_{1} \vee V_{2}.\]
  
\noindent Since $(\overline{V}_{1})^{ls} \cap (\overline{V}_{2})^{ls}=\emptyset$  and  $c_{\overline{V}_{1}} < c_{\overline{V}_{2}}$, the property TCLS-AU ensures that  $F(N,W,P) = F(N,V_{1},P).$\medskip
  
\noindent Consider now the collection $\bigl\{(N,W^{T_{h}},P)\mid\hspace{1mm} h \in \{1, \ldots, l'\}\bigr\}$. For each  $h\in \{1,\ldots, l'\}$, it follows that $(\overline{W^{T_{h}}})^{m}=(\overline{W^{T_{h}}})^{ls}= \{R_{h}\}$, $p_{\hspace{1mm}\overline{W^{T_{h}}}}=1$ and $c_{\hspace{1mm}\overline{W^{T_{h}}}}=c_{\hspace{1mm}\overline{W}}=| R_{h}|$. Additionally, for any $ h_{1},h_{2} \in \{1, \ldots, l'\}$  we have that $(\overline{W^{T_{h_{1}}}})^{ls}\cap (\overline{W^{T_{h_{2}}}})^{ls}= \{R_{h_{1}}\}\cap \{R_{h_{2}}\}= \emptyset$ and $c_{\hspace{1mm}\overline{W^{T_{h_{1}}}}}= c_{\hspace{1mm}\overline{W^{T_{h_{2}}}}}$ for $h_{1}\neq h_{2}.$ Applying TCLS-AU property again and noting that $ \overline{W}^{ls}= \overline{V_{1}}^{ls}$, we obtain:
\begin{equation}\label{eq:charac2_unic_5}
   F(N,V_{1},P) =\frac{1}{\big| \overline{V_{1}}^{ls}\big| }\sum_{h=1}^{l'} p_{\hspace{0.5mm}\overline{W^{T_{h}}}}\hspace{1mm}F(N,W^{T_{h}},P) 
        = \frac{1}{\big| \overline{W}^{ls}\big| }\sum_{h=1}^{l'}  \hspace{0.5mm}F(N,W^{T_{h}},P).
    \end{equation}
    
\noindent Next, to analyze $F(N,W^{T_{h}},P)$, we define the disjunction game $W_{h}=\vee_{k\in R_{h}} W_{R_{h},k}$, where each component is determined by $(W_{R_{h},k})^{m}= E_{R_{h},k}(N,W^{T_h},P).$\\
Since $E^{ls}(N,W_{h},P)= E^{ls}(N,W^{T_{h}},P)$ and 
$(\overline{W^{T_{h}}})^{ls} = \{R_{h}\}$, whereas $(\overline{W_{h}})^{ls} = \{\{k\} \mid {k\in R_{h}}\}$, the ILSE property ensures that $F(N,W^{T_{h}},P)=F(N,W_{h},P)$.\medskip

\noindent We now $F(N,W_{h},P)$ decompose by reapplying once more the TCLS-AU property. Consider the collection  $\{(N,W_{R_{h},k},P) \mid k\in R_{h}\}$ in which for any $k_{1},k_{2}\in R_{h}$ we have that $(\overline{W_{R_{h},k_{1}}})^{ls}\cap(\overline{W_{R_{h},k_{2}}})^{ls}= \{\{k_{1}\}\}\cap\{\{k_{2}\}\}=  \emptyset$ and $c_{\hspace{1mm}\overline{W_{R_{h},k_{1}}}} = c_{\hspace{1mm}\overline{W_{R_{h},k_{2}}}}= 1$, $k_{1}\neq k_{2}$. Given a player $i\in P_{k}$, there are two options:
\begin{enumerate}
    \item If  $k\notin R_{h}$, then $i$ is a null player in the game $(N,W_{h},P)$ and the NP property implies that $F_{i}(N,W_{h},P)=0$.
    \item Otherwise,  $k\in R_{h}$ and by TCLS-AU:
    \begin{equation}\label{eq:charac2_unic_6}
          \begin{aligned}[t]
     F_{i}(N,W_{h},P) &= \frac{1}{p_{\hspace{0.4mm}\overline{W_{h}}}}\sum_{\tilde{k}\in R_{h}}p_{\hspace{1mm}\overline{W_{R_{h},\tilde{k}}}}F_{i}(N,W_{R_{h},\tilde{k}},P) \\
        &= \frac{1}{|R_{h}|}\sum_{\tilde{k}\in R_{h}} F_{i}(N,W_{R_{h},\tilde{k}},P) \\
               &= \frac{1}{|R_{h}|}F_{i}(N,W_{R_{h},k} ,P),
  \end{aligned}
    \end{equation}
     where the equalities follow from the fact that $p_{\hspace{1mm}\overline{W_{R_{h},\tilde{k}}}}=1$ for all $ \tilde{k}\in R_{h}$, and that $i\in P_{k}$ is a null player in each game $(N,W_{R_{h},\tilde{k}},P)$ with $\tilde{k}\in R_{h}$, $\tilde{k} \neq k$ combined with the assumption that $F$ satisfies the NP property.\medskip
     
    \noindent Lastly, $F_{i}(N,W_{R_{h},k} ,P)$ is decomposed by considering the TCLS-IU property.   Note that $W_{R_{h},k}=W_{L_{1}} \vee \ldots \vee W_{L_{z}}$, with $L_{j} \subseteq P_{k}$ and $(N,W_{L_{j}},P)$ the unanimity game of $L_{j}$. Consider the collection  $\{(N,W_{L_{j}},P) \mid j\in \{1,\ldots, z\}\} $,  for any $j_{1},j_{2}\in \{1,\ldots, z\} $,   $(W_{L_{j_{1}}})^{ls}\cap (W_{L_{j_{2}}})^{ls}= \emptyset$ and 
$c_{\hspace{1mm}\overline{W_{L_{j_{1}}}}} = c_{\hspace{1mm}\overline{W_{L_{j_{2}}}}}$ for $j_{1}\neq j_{2}$. Thus, TCLS-IU implies that:
\begin{equation}\label{eq:charac2_unic_7}
          \begin{aligned}[t]
      F_{i}(N,W_{R_{h},k},P)  &=\frac{1}{p_{\hspace{0.5mm}W_{R_{h},k}}}\sum_{j=1}^{z}p_{\hspace{0.5mm}W_{L_{j}}} \hspace{0.5mm}F_{i}(N,W_{L_{j}},P) \\
        &= \frac{1}{|E_{R_{h},k}^{ls}(N,W,P)|}\sum_{S\in E_{R_{h},k,i}^{ls}(N,W,P)} \hspace{0.5mm}F_{i}(N,W_{S},P).
  \end{aligned}
    \end{equation}
     where the last equality follows from  the fact that $i$ is a null player in every game $(N,W_{L_{j}},P)$ such that $i \notin L_{j}$ and $F$ satisfies the NP property.
\end{enumerate}
\noindent Taken together the results~(\ref{eq:charac2_unic_1}), (\ref{eq:charac2_unic_5}), (\ref{eq:charac2_unic_6}), and (\ref{eq:charac2_unic_7}), we have that for $i\in P_{k}$,
\begin{align*}
      F_{i}(N,W,P) & =\frac{1}{\big| \overline{W}^{ls}\big| }\sum_{h=1}^{l'}  \hspace{0.5mm}F_{i}(N,W^{T_{h}},P)&\\
      & =  \frac{1}{\big| \overline{W}^{ls}\big| }\sum_{h=1}^{l'}  \hspace{0.5mm} \frac{1}{|R_{h}|}F_{i}(N,W_{R_{h},k},P)&\\
      & =  \frac{1}{\big| \overline{W}^{ls}\big| }\sum_{h=1}^{l'}  \hspace{0.5mm}    \frac{1}{|R_{h}|}\frac{1}{|E_{R_{h},k}^{ls}(N,W,P)|}\sum_{S\in E_{R_{h},k,i}^{ls}(N,W,P)} \hspace{0.5mm}F_{i}(N,W_{S},P)&\\
      & =  \frac{1}{|\overline{W}^{ls}|}\sum_{R\in \overline{W}^{ls}_{k}}\frac{1}{|R|}\frac{1}{|E_{R,k}^{ls}(N,W,P)|} \sum_{S\in E_{R,k,i}^{ls}(N,W,P)}\frac{1}{|S|}.
    \end{align*} 
    Thus, we conclude that  $F$ is the Felsenthal Owen  power index $\Psi$.
    
\end{proof}

\section{Application example: The International Monetary Fund}

The International Monetary Fund (IMF) is a global financial organization engaged  with the economic well-being and development of its 191 member countries. Among its main objectives, the IMF promotes international monetary cooperation, fosters global trade expansion, and provides temporary financial assistance to countries facing economic difficulties. 


The highest decision-making body of the IMF is the \textit{Board of Governors}, consisting at present of 191 governors, each representing a member country.  Governors are assigned voting weights based on their country's financial contribution to the IMF, and decisions are made through a weighted voting system. The required quota for reaching a decision varies depending on its importance, with thresholds of 50\%, 70\%, or 85\%.

In practice, the Board of Governors normally convenes once a year. To enhance the efficiency of IMF's activity, a smaller body known as the \textit{Executive Board}, is entrusted with the task of daily decision-making.
The structure of this smaller board is the result of negotiations among member countries, which organize to form groups. Once the groups are formed, each one chooses a representative that casts their votes in the Executive Board, on behalf of the entire group. The groups are called the \textit{constituencies} and the representatives are the \textit{executive directors}. Currently, the body consists of 25 executive directors and, as in the Board of Governors, decisions require a specific percentage of weighted votes (50\%, 70\%, or 85\%).
In this section, we analyze the distribution of power among IMF members based on the governance structure as of March 2025 \footnote{The current structure can be inspected on the  \href{https://www.imf.org/en/About/executive-board/eds-voting-power}{IMF's website}.}. This analysis is of great pertinence, as the IMF frequently requests guidance to decide weight realignments. It is worth mentioning that, at the moment, IMF’s dealings with 3 member countries --Afghanistan, Myanmar, and Venezuela-- are paused due to government recognition issues. In consequence, they are neither represented at the Executive Board nor included in our study.

Based on the described governance structures, decision-making in both the Board of Governors and the Executive Board can be modelled as simple games; $(N,W)$  for the Board of governors with 188 players and $(U,\overline{W})$ for the Executive Board with 25 players. Additionally, if we consider the partition $P$ of the IMF countries into constituencies,  the Executive Board game corresponds precisely to the quotient game of the game with a priori unions $(N,W,P)$.

\begin{figure}[H]
    \centering
    \includegraphics[width=0.75\textwidth]{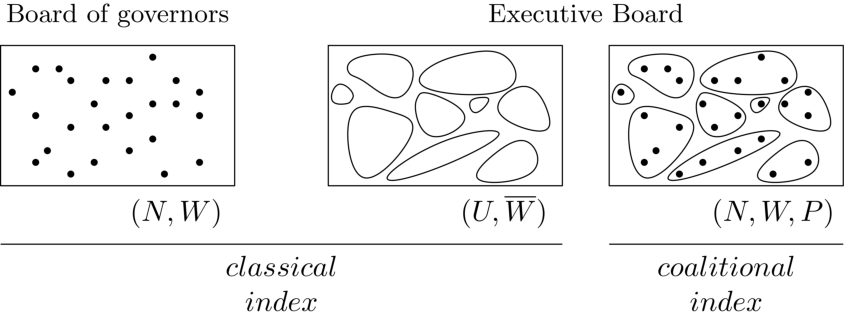}
    \caption{IMF structure and appropriate indexes.}
    \label{fig:estructura_IMF} 
\end{figure}

Regarding the assessment of power, classical (non-coalitional) power indices are suitable for measuring the influence exerted by each country $i\in N$ within the Board of Governors $(N,W)$, as shown in previous studies
(see \cite{49_livino_IMF} and \cite{67_alex_saavedra_IMF}). Similarly, they can also be used to determine power distribution among constituencies $k\in U$ within the Executive Board $(U,\overline{W})$. However, classical indexes do not capture the individual power of countries  within a constituency in the Executive Board, $i\in P_{k}$. 
This limitation can be addressed by coalitional indexes, which provide a valuable solution by evaluating individual power while considering constituencies as a priori coalitions, as emphasized by Alonso-Meijide and Bowles in \cite{14_generating_functions_IMF}. Figure~\ref{fig:estructura_IMF} illustrates the structure of the IMF and the appropriate power indices for each governing body.
Considering these precedents, we have calculated the \textit{Felsenthal power index} and the \textit{Felsenthal Owen power index}, to reveal the power of each country in the Board of Governors and Executive Board, respectively.  These indices allow us to assess how a country’s influence evolves as decision-making shifts from individual voting weights in the Board of Governors to collective representation through constituencies in the Executive Board.
 The complete results for both indices are provided in the ORS. Below, we provide a general overview of the findings as of March 2025, under a 50\% quota. 



\begin{figure}[H]
    \centering
    \includegraphics[width=0.97\textwidth]{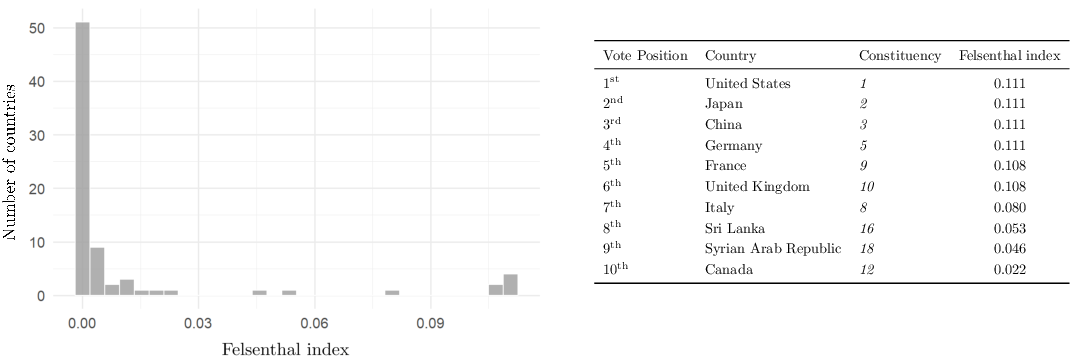}
    \caption{  On the left, the histogram of the non-zero values of the Felsenthal index; on the right, the list of ten Countries with the highest Felsenthal index, in descending order.}
    \label{fig:felsenthal_histograma_tabla}
\end{figure}

\begin{figure}[H]
    \centering
    \includegraphics[width=0.95\textwidth]{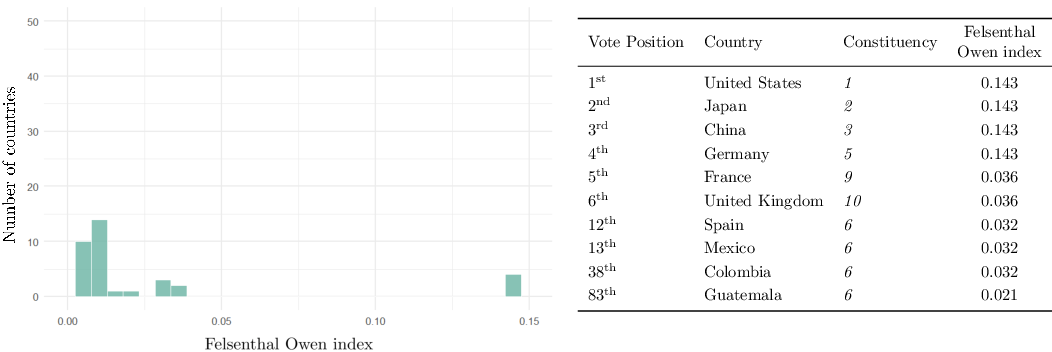}
    \caption{  On the left, the histogram of the non-zero values of the Felsenthal Owen index; on the right, the list of ten Countries with the highest Felsenthal Owen index, in descending order. }
    \label{fig:felsenthal_owen_histograma_tabla}
\end{figure}

 \textbf{Felsenthal index.} Table 1 of the ORS presents the Felsenthal index values for each country represented in the Board of Governors. First of all, it is important to note that the minimum size of a winning coalition is 9, meaning that at least 9 countries must agree to approve a decision in the Board of Governors ($c_{W}= 9$). Focusing on power distribution, Table 1 shows that only 77 out of the 188 countries have a Felsenthal power index greater than 0. Figure~\ref{fig:felsenthal_histograma_tabla} displays a histogram of these 77 non-zero values coloured in gray. The histogram reveals a notable imbalance in power distribution even among non-null members, with most countries displaying values close to zero while a few concentrate the majority of power. In fact, the six most powerful countries according to this index together account for 66\% of the total power, as shown in the table of Figure~\ref{fig:felsenthal_histograma_tabla}, which lists the ten countries with the highest Felsenthal power index in decreasing order.
\textbf{Felsenthal Owen index.} In the case of the Executive Board, only seven constituencies are required to form a winning coalition ($c_{\overline{W}}= 7$), consistent with the body's design to enhance the efficiency of the Board of Governors. The Felsenthal Owen index results, detailed in Table 1 of the ORS, 
 indicate a further concentration of power in the Executive Board: only 38 countries exhibit a non-zero index. 
The distribution of these values, as illustrated in the histogram of Figure~\ref{fig:felsenthal_owen_histograma_tabla}, mirrors the asymmetry observed within the Board of Governors, with a small group of countries dominating decision-making. In fact, the six with the highest Felsenthal Owen index values coincide with those leading the Felsenthal index ranking.
In this case, they collectively hold 64.3\% of the total power, which can be observed in the table of Figure~\ref{fig:felsenthal_owen_histograma_tabla} sorted in descending order according to the Felsenthal Owen index. Despite these similarities, the formation of constituencies gives rise to differences in the allocation of power between those six leading countries. While their Felsenthal indexes are close (around 0.1), the Felsenthal Owen index delineates a pronounced gap: the leading four countries display values nearly four times greater than those of France and the UK.

\begin{figure}[H]
    \centering
    \includegraphics[width=0.8\textwidth]{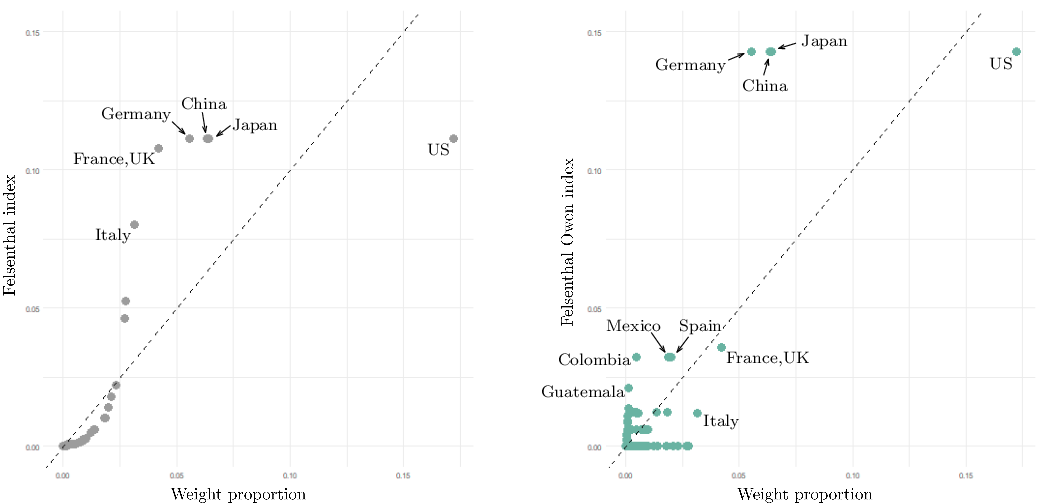}
 \caption{Scatterplots showing the proportion of weight over the total for different countries on the X-axis, and the corresponding values Felsenthal and Felsenthal Owen indexes on the Y-axis (left and right, respectively).}
    \label{fig:peso_felsenthal_felsenthal_owen}
\end{figure}

Beyond examining power distribution within each Board, it is essential to explore the relationship between a country's voting weight and its actual voting power, given the frequent discrepancies between both \footnote{ A similar analysis can be found in \cite{77_IMF_weight_vs_power}.}. A comparison is carried out using the two scatterplots presented in Figure~\ref{fig:peso_felsenthal_felsenthal_owen}.

\textbf{Felsenthal index.} The scatterplot on the left illustrates the relationship between the proportion of total weight and the Felsenthal index. The plot suggests a positive correlation between power and weight, with higher values of one variable being associated with higher values of the other. This direct relationship was already evident in the table of Figure~\ref{fig:felsenthal_histograma_tabla}, where the order given by the Felsenthal index corresponded with the weight rank shown in the first column. However, the plot also reveals that this relationship is far from proportional, as the data points considerably deviate from the (dashed) identity line. Notably, countries with a weight proportion exceeding 0.025, such as France, the UK, and Italy, possess significantly  more power than their voting weights would suggest. The exception is the United States,  whose power aligns with the other three countries with greater weight (Japan, China, and Germany), despite contributing significantly more weight itself.

\textbf{Felsenthal Owen index.} The scatterplot on the right corresponds to the Felsenthal Owen index. It reveals that the formation of constituencies significantly disrupts the proportionality between power and weight. For instance, countries with a light voting share such as Spain, Mexico, Colombia, and Guatemala, benefit from constituency formation, achieving a strategic position within the Executive Board. This was already noticeable in the table of Figure~\ref{fig:felsenthal_owen_histograma_tabla}, where these countries rank among the 10 leading countries despite their lower weight rankings (12\textsuperscript{th}, 13\textsuperscript{th}, 38\textsuperscript{th}, and 83\textsuperscript{th}, respectively). Conversely,  countries such as France, United Kingdom, and Italy lose significant power due to grouping. Finally, just mention that the four countries with greater weight, become even more dominant in the Executive Board, as previously noted in the histogram analysis. 

These results suggest that, based on the weights and constituencies as of March 2025 and a quota of 50\%, power is unevenly distributed among countries in both the Board of Governors and the Executive Board, with the same six countries—the ones with greater weight—dominating both bodies. This imbalance is even more pronounced in the Executive Board, where the number of null players increases and power is more concentrated in the first four countries, consolidating their influence. The remaining countries are also affected by the formation of constituencies, as these groupings can enhance the position of some lower-weight countries while reducing the power of others with greater weight. 

\begin{figure}[H]
\centering
    \includegraphics[width=0.8\textwidth]{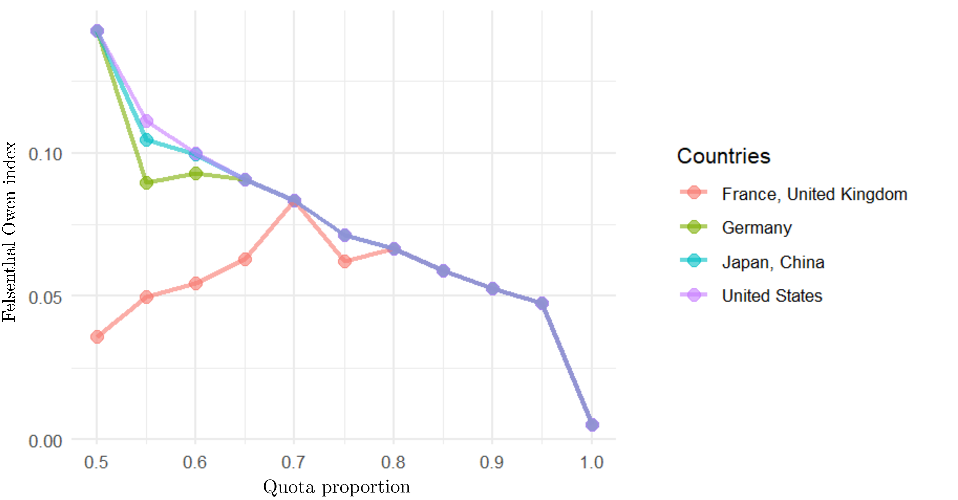}

\caption{The effect of the majority requirement on the value of the Felsenthal Owen index of the six countries with greater weight. }
    \label{fig:felsenthalowen_top6(quotas)}
\end{figure}

The present analysis has been focused on the majority requirement for ordinary decisions $q=50\%$. However, higher thresholds such as those required for more significant decisions ($q=70\%, q=85\%$), can lead to substantial changes in power distribution, and it is worth to undertake further exploration on how increases in the quota affects power allocation. Such an analysis could offer valuable insights for determining the appropriate quota for a specific decision, ensuring that power is distributed among members as intended \footnote{A similar approach can be found in \cite{60_Leech_IMF}.} .\\[2pt] 
Figure~\ref{fig:felsenthalowen_top6(quotas)} and illustrates the impact of quota variations on power within the Executive Board for the six leading countries: United States, Japan, China, Germany, France, and United Kingdom. The plot was generated by calculating the Felsenthal Owen power index for quota values ranging from 0.5 to 1, with increments 0.05. Each dominant country is represented by a distinct line on the plot.  It should be mentioned that both China and Japan, as well as France and the UK, exhibit identical values of the Felsenthal Owen indices across the proposed quotas and therefore each pair is represented by the same line.

The plot suggests that, in general terms, an increase of the majority requirement tends to equalise power. As the quota rises, dominant countries experience a decline in their influence approaching the limit $q=100\%$, that correspond to a unanimity voting rule under which all participating countries hold equal same power. The exceptions are France and the UK, with an increase in power up to 70\%.  However, beyond 80\%, the strategic positions of all countries converge, following a descending trend.\\[7pt]

\section*{Conclusion}

Assessing the influence of participants in a decision-making context is often based on voting weights, a practice which frequently leads to erroneous conclusions. Herein, we have demonstrated that modelling decision scenarios within the framework of cooperative game theory, and using a novel power index, the \textit{Felsenthal Owen index}, provides a more accurate measure of each player’s actual influence. This \textit{Felsenthal Owen index} is particularly valuable for two primary reasons. First, the index is built on the realistic assumption that only winning coalitions of least size are feasible, reflecting the individual ambition and strategic behaviour of players. Second, it incorporates a priori unions, capturing possible pre-grouping of players.

Moreover, two distinct axiomatic characterizations of the new index are provided. These not only deepen our understanding of the features of the index, but also facilitate comparison with other existing power indexes, as many of them share several properties with ours. It is important to reiterate the absence of a universally applicable dogmatic power index, given that each index is deemed suitable under a particular decision scenario and a designated purpose.

We have applied the Felsenthal Owen index to examine the allocation of power among the countries in the International Monetary Fund, an institution whose structure fits perfectly within our analytical framework. Three major findings emerge from our results. First, power distribution shifts depending on how countries participate in decision-making processes, whether as individual entities or as part of a constituency formed through the grouping of countries. Second, we identify significant discrepancies between a member’s voting weight and its actual influence. Lastly, the distribution of power proves highly sensitive to the majority threshold fixed for the voting. 
We encourage to take these results into account (or recalculate the index if necessary) in future IMF reforms, such as the reassignment of voting weights, the restructuring of constituencies, or the redefinition of majority requirements. Namely, our power index can shed light on the paradoxical behaviour of power and help ensure that institutional arrangements align better with their intended purposes.

It is essential to acknowledge the limitations of the proposed power index to ensure its proper use and to draw reliable conclusions from its application. Decision-making processes are inherently complex and often influenced by unwritten factors that become relevant in the power dynamics, such as informal consensus. Some of those factors are either unknown or difficult to incorporate into a mathematical model.
For this reason, power indexes should not be regarded as absolute measures of power, yet far more informative than voting weights.

That said, the \textit{Felsenthal Owen power index} may be further polished to better capture the nuances of real decision-making scenarios. For instance, the possibility of abstention could be incorporated into the model, as proposed in \cite{83_felsenthal_ternary_abstention}. Furthermore, rather than assuming that all least size winning coalitions are equally probable, it is possible to account for the fact that some coalitions may be more likely than others, or even infeasible, following the approach outlined in \cite{82_Restriction_carreras}.

\section*{Acknowledgements}
This publication is part of the R+D+I project grants PID2021-12403030NB-C31 and  PID2021-124030NB-C32 funded by MCIN/AEI/10.13039/501100011033/ and by ERDF/EU. It has also been supported by the Xunta de Galicia (Grupos de Referencia Competitiva ED431C-2024/14) and by CITIC as a center accredited for excellence within the Galician University System and a member of the CIGUS Network, receives subsidies from the Department of Education, Science, Universities, and Vocational Training of the Xunta de Galicia. Additionally, it is co-financed by the EU through the FEDER Galicia 2021-27 operational program (Ref. ED431G 2023/01).  The first author received funding under the Spanish Predoctoral Research Trainees Program (RD 103/2019), as part of grant PRE2022-104510, funded by MCIN/AEI/10.13039/501100011033/ and ESF+.
\section*{Online Resource Section}
The Online Resource Section (ORS) accompanying this paper provides the full computations of the Felsenthal index and the Felsenthal Owen index, applied to the IMF governance structure as of March 2025, for decision thresholds of 50\%, 70\%, and 80\%. These results form the basis for the analysis presented in Section 4.

\section*{Appendix}

\begin{rem}
All the properties in Theorem~\ref{Theorem1} are independent.
\begin{itemize}
    \item The coalitional power index defined for each $(N,W,P)\in SIU(N)$ and each $i\in N$ as $F^1_{i}(N,W,P)=0$, satisfies NN, QG, and PELS except CFI.
    \item For any $S\subseteq N$, $S\neq \emptyset$, let us consider the minimum and maximum members of $S$ according to the ordering on the set of natural numbers. Given $(N,W,P)\in SIU(N)$, let us define the coalitional index $F^2$ for a player $i\in P_{k}$ by distinguishing two situations:
\begin{itemize}
    \item [(i)] If $|P_{k}|>1$ and all $j \in P_{k}$ are null in $(N,W)$, then
 
    \begin{equation*}F^2_{i}(N,W,P)=\left \{
    \begin{array}{rl}
      -1, & \text{ if $i$ is the minimum of } P_{k} \\
     1, &  \text{ if $i$ is the maximum of } P_{k}  \\
      0, &  \text{ otherwise.}  
   \end{array}\right.\end{equation*}
    
    \item[(ii)] Otherwise, $F^2_{i}(N,W,P)=\Psi_{i}(N,W,P).$
\end{itemize}
It can be proved that $F^2$ satisfies all the properties except NN.

\item Given $(N,W,P)\in SIU(N)$, the coalitional index $F^3$ is defined for each $i\in P_{k}$ according to the following conditions:
\begin{itemize}
    \item [(i)] If $\overline{W}^{ls}_{k}= \emptyset$, then $F^3_{i}(N,W,P)=\psi_{i}(N,W)$.
    \item[(ii)] Otherwise, $F^3_{i}(N,W,P)=\Psi_{i}(N,W,P)$.
\end{itemize}
It is straightforward to prove that $F^3$ satisfies all the properties except QG.

\item Given $(N,W,P)\in SIU(N)$, we define the coalitional index $F^4$ for a player $i\in P_{k}$ as
\begin{equation*} 
F^4_{i}(N,W,P) =\frac{1}{|\overline{W}^{ls}|}\sum_{R\in \overline{W}^{ls}_{k}}\frac{1}{|R|}\frac{1}{|E^{m}_{R,k}(N,W,P)|} \sum_{S\in E_{R,k,i}(N,W,P)}\frac{1}{|S|}.
\end{equation*}
Following a similar procedure to the Felsenthal Owen power index, it can be shown that $F^4$ satisfies all the properties except PELS.
\end{itemize}
\end{rem}

\begin{rem}
All the properties in Theorem~\ref{Theorem2} are independent.
\begin{itemize}
    \item The coalitional power index $F^1$ satisfies all the properties except E.
    
    \item For each $(N,W,P)\in SIU(N)$, let us consider for each $i\in P_k\in P$, 
$$
F^5_{i}(N,W,P)=\frac{1}{|U|}\frac{1}{|P_{k}|}.$$
The coalitional power index $F^5$ satisfies all the properties except NP.

\item Given $(N,W,P)\in SIU(N)$, let us define for all $i\in N$ the coalitional index $F^6$ as $F^6_{i}(N,W,P)=$
\begin{equation*}
\left\{
    \begin{array}{lr}
        \dfrac{i \Psi_{i}(N,W,N^{0})}{\sum\limits_{h\in N}h \Psi_{h}(N,W,N^{0})}, & \text{if } W=\{N\} \text{ or }  W^{m}=\{\{1\},\ldots,\{n\}\} \\[25pt]
        \Psi_{i}(N,W,P), & \text{otherwise.}
    \end{array}
\right.
\end{equation*}
The coalitional power index $F^6$ satisfies all the properties except S-AU.

\item  Given $(N,W,P)\in SIU(N)$ and $i\in N$, let us define the coalitional index  $F^7_{i}(N,W,P)=\Psi_{i}(N,W,P)$ for all $(N,W,P) \in SIU(N)$ except for the game $(N,\{N\},N^{N})$, where the index is defined as
$$
 F^7_{i}(N,\{N\},N^{N})=\dfrac{i \Psi_{i}(N,\{N\},N^{N})}{\sum\limits_{h\in N}h \Psi_{h}(N,\{N\},N^{N})}.
$$


The coalitional power index $F^7$ satisfies all the properties except S-IU. 

\item Given $(N,W,P)\in SIU(N)$ and $i\in P_k\in P$, let us define the coalitional index $F^{8}_{i}(N,W,P)=\Psi_{i}(N,W,P)$ except for the game $(\widetilde{N},\widetilde{W},\widetilde{P})$ with 
$\widetilde{N}=\{1,2,3,4,5\}$, $\widetilde{P}=\{\{1\},\{2,3,4\},\{5\}\}$, and  $(\widetilde{W})^{m}=\{\{1,5\},\{1,2,3,4\}\}$, where the index is defined as 
$$
 F^{8}_{i}(\widetilde{N},\widetilde{W},\widetilde{P}) =\phi_{k}(\widetilde{U},\overline{\widetilde{W}})\dfrac{ \Psi_{i}(\widetilde{N},\widetilde{W},\widetilde{P})}{\sum\limits_{h\in P_{k}} \Psi_{h}(\widetilde{N},\widetilde{W},\widetilde{P})},
$$
 with $\phi$ denoting the Shapley-Shubik power index~\cite{shapley_shubik}.The coalitional power index $F^8$ satisfies all the properties except TCLS-AU. \medskip

\item Let us define the coalitional power index  $F^{9}(N,W,P)=\Psi(N,W,P)$ for all $(N,W,P) \in SIU(N)$ except for the case in which $\widetilde{N}=\{1,2,3\}$ and $P=\{\widetilde{N}\}$, where the index is defined as
     \begin{equation*}
F^{9}_{i}(\widetilde{N},W,\{\widetilde{N}\})=
\left\{
    \begin{array}{lr}
        0, & \text{if } W^{m}_{i}=\emptyset \\
        \dfrac{1}{|ls(\widetilde{N},W)|}, & \text{ otherwise, }
    \end{array}
\right.
\end{equation*}
where $ls(\widetilde{N},W)$ denotes the set of players $i\in \widetilde{N}$ such that $W^{ls}_{i}\neq \emptyset$. The coalitional power index $F^9$ satisfies all the properties except TCLS-IU.

 \item Given $(N,W,P)\in SIU(N)$, 
let us define $F^{10}(N,W,P)=\Psi(N,W,P)$ for all $(N,W,P) \in SIU(N)$ except for the case in which $N=\{1,2,3,4,5\}$, $W^{m}=\{\{1,5\},\{1,2,4\},\{2,3,5\},\{1,2,5\}\}$, and $P=\{\{1\},\{2,3,4\},\{5\}\}$. In this case, for each $i\in P_{k}\in P$,  $F^{10}$ is given by
\begin{equation*}
 F^{10}_{i}(N,W,P)=
\left\{
    \begin{array}{lr}
       \Psi_i(N,W,P), &  i = 1,2,5\\[5pt]
       \Psi_i(N,W,P) -\epsilon, &  i = 3\\[5pt]
       \Psi_i(N,W,P) + \epsilon, &  i = 4,
    \end{array}
\right.
\end{equation*}
assuming that $\epsilon \approx 0$. The coalitional power index $F^{10}$ satisfies all the properties except IIC.

\item Given $(N,W,P)\in SIU(N)$, the coalitional power index $F^{11}$ is defined for a player $i\in P_{k}\in P$ as
$$
F^{11}_{i}(N,W,P)= \frac{1}{|\overline{W}^{ls}|}\sum_{R\in \overline{W}^{ls}_{k}}\frac{1}{|R|}\frac{1}{|C_{R,k}|}\sum_{S\in C_{R,k,i}}\frac{1}{|S|},
$$
where $C_{R,k}=\{S\cap P_{k} \mid S \in W^{m} \text{ and } u(S)=R \}$ and $C_{R,k,i}=\{S\in C_{R,k} \mid i\in S\}$. 
The coalitional power index $F^{11}$ satisfies all the properties except ILSE. \medskip

\end{itemize}
\end{rem}

\bibliographystyle{abbrv}  
\bibliography{bibliografia}

\end{document}


\maketitle
\vspace{-3em}
\noindent Table 1, included in this ORS, presents the values of the Felsenthal index and the Felsenthal Owen index for each member country of the IMF, based on the voting weights and constituency structure as of March 2025. Results are provided for decision thresholds of 50\%, 70\%, and 80\%. Countries are ranked in descending order by the total voting weight of their constituency, and within each constituency, by individual voting weight.

\renewcommand{\arraystretch}{1.5} 
\begin{table}[htbp]
\centering
\caption{Voting Power at the IMF in March 2025}
\begin{adjustbox}{width=0.95\textwidth, totalheight=\textheight, keepaspectratio}
\begin{tabular}{lcccccccc}
\toprule
\multirow[b]{2}{*}{\textbf{Country}} & \multirow[b]{2}{*}{\makecell{\textbf{Voting}\\\textbf{weight}}} & \multirow[b]{2}{*}{\textbf{Constituency}} &
\multicolumn{3}{c}{\textbf{Felsenthal index}} &
\multicolumn{3}{c}{\textbf{Felsenthal Owen index}} \\
\cmidrule(lr){4-6} \cmidrule(lr){7-9}
& & & $q=50\%$ & $q=70\%$ & $q=85\%$ & $q=50\%$ & $q=70\%$ & $q=85\%$ \\[0.2em]
\midrule
United States & 831394 & 1 & 0.1111 & 0.0500 & 0.0227 & 0.1429 & 0.0833 & 0.0588 \\ 
\hdashline
Japan & 309657 & 2 & 0.1111 & 0.0500 & 0.0227 & 0.1429 & 0.0833 & 0.0588 \\
\hdashline
China & 306281 & 3 & 0.1111 & 0.0500 & 0.0227 & 0.1429 & 0.0833 & 0.0588 \\
\hdashline
Netherlands & 88817 & 4 & 0.0101 & 0.0500 & 0.0227 & 0.0123 & 0.0076 & 0.0059 \\

Belgium & 65559 & 4 & 0.0061 & 0.0489 & 0.0227 & 0.0123 & 0.0076 & 0.0059 \\

Ukraine & 21570 & 4 & 0.0008 & 0.0000 & 0.0155 & 0.0123 & 0.0076 & 0.0058 \\

Israel & 20661 & 4 & 0.0008 & 0.0000 & 0.0137 & 0.0123 & 0.0076 & 0.0058 \\

Romania & 19566 & 4 & 0.0008 & 0.0000 & 0.0113 & 0.0123 & 0.0076 & 0.0058 \\

Luxembourg & 14670 & 4 & 0.0008 & 0.0000 & 0.0031 & 0.0123 & 0.0076 & 0.0055 \\

Bulgaria & 10415 & 4 & 0.0008 & 0.0000 & 0.0006 & 0.0123 & 0.0076 & 0.0046 \\

Croatia & 8626 & 4 & 0.0008 & 0.0000 & 0.0003 & 0.0123 & 0.0076 & 0.0042 \\

Cyprus & 4490 & 4 & 0.0000 & 0.0000 & 0.0000 & 0.0109 & 0.0055 & 0.0026 \\

    Bosnia and Herzegovina & 4104  & 4     & 0.0000 & 0.0000 & 0.0000 & 0.0084 & 0.0036 & 0.0024 \\
    Georgia & 3556  & 4     & 0.0000 & 0.0000 & 0.0000 & 0.0061 & 0.0034 & 0.0023 \\
    Moldova & 3177  & 4     & 0.0000 & 0.0000 & 0.0000 & 0.0059 & 0.0029 & 0.0019 \\
    North Macedonia & 2855  & 4     & 0.0000 & 0.0000 & 0.0000 & 0.0043 & 0.0022 & 0.0017 \\
    Armenia & 2740  & 4     & 0.0000 & 0.0000 & 0.0000 & 0.0041 & 0.0022 & 0.0017 \\
    Andorra & 2277  & 4     & 0.0000 & 0.0000 & 0.0000 & 0.0025 & 0.0014 & 0.0015 \\
    Montenegro & 2057  & 4     & 0.0000 & 0.0000 & 0.0000 & 0.0021 & 0.0014 & 0.0015 \\
    \hdashline
    Germany & 267796 & 5     & 0.1111 & 0.0500 & 0.0227 & 0.1429 & 0.0833 & 0.0588 \\
    \hdashline
    
    Spain & 96807 & 6     & 0.0142 & 0.0500 & 0.0227 & 0.0321 & 0.0190 & 0.0157 \\
    Mexico & 90579 & 6     & 0.0101 & 0.0500 & 0.0227 & 0.0321 & 0.0190 & 0.0157 \\
    Colombia & 21897 & 6     & 0.0008 & 0.0000 & 0.0161 & 0.0321 & 0.0190 & 0.0111 \\
    Guatemala & 5738  & 6     & 0.0000 & 0.0000 & 0.0001 & 0.0210 & 0.0066 & 0.0044 \\
    Costa Rica & 5146  & 6     & 0.0000 & 0.0000 & 0.0000 & 0.0138 & 0.0066 & 0.0042 \\
    El Salvador & 4324  & 6     & 0.0000 & 0.0000 & 0.0000 & 0.0094 & 0.0066 & 0.0039 \\
    Honduras & 3950  & 6     & 0.0000 & 0.0000 & 0.0000 & 0.0022 & 0.0066 & 0.0037 \\
\bottomrule
\end{tabular}%
\end{adjustbox}
\end{table}

\begin{table}[htbp]
\centering
\caption*{Table 1: Voting Power at the IMF in March 2025 (\textit{Continued})}
\begin{adjustbox}{width=0.95\textwidth, totalheight=\textheight, keepaspectratio}
\begin{tabular}{lcccccccc}
\toprule
\multirow[b]{2}{*}{\textbf{Country}} & \multirow[b]{2}{*}{\makecell{\textbf{Voting}\\\textbf{weight}}} & \multirow[b]{2}{*}{\textbf{Constituency}} &
\multicolumn{3}{c}{\textbf{Felsenthal index}} &
\multicolumn{3}{c}{\textbf{Felsenthal Owen index}} \\
\cmidrule(lr){4-6} \cmidrule(lr){7-9}
& & & $q=50\%$ & $q=70\%$ & $q=85\%$ & $q=50\%$ & $q=70\%$ & $q=85\%$ \\[0.2em]
\midrule
    Indonesia & 47936 & 7     & 0.0028 & 0.0168 & 0.0227 & 0.0060 & 0.0110 & 0.0086 \\
    Singapore & 40371 & 7     & 0.0020 & 0.0079 & 0.0227 & 0.0060 & 0.0110 & 0.0086 \\
    Malaysia & 37790 & 7     & 0.0020 & 0.0058 & 0.0227 & 0.0060 & 0.0110 & 0.0086 \\
    Thailand & 33571 & 7     & 0.0016 & 0.0037 & 0.0227 & 0.0060 & 0.0110 & 0.0084 \\
    Philippines & 21881 & 7     & 0.0008 & 0.0000 & 0.0161 & 0.0060 & 0.0110 & 0.0075 \\
    Vietnam & 12983 & 7     & 0.0008 & 0.0000 & 0.0018 & 0.0060 & 0.0110 & 0.0054 \\
    Brunei Darussalam & 4465  & 7     & 0.0000 & 0.0000 & 0.0000 & 0.0000 & 0.0041 & 0.0029 \\
    Cambodia & 3202  & 7     & 0.0000 & 0.0000 & 0.0000 & 0.0000 & 0.0030 & 0.0022 \\
    Nepal & 3021  & 7     & 0.0000 & 0.0000 & 0.0000 & 0.0000 & 0.0030 & 0.0021 \\
    Lao People's Democratic Republic & 2510  & 7     & 0.0000 & 0.0000 & 0.0000 & 0.0000 & 0.0030 & 0.0016 \\
    Fiji, Republic of & 2436  & 7     & 0.0000 & 0.0000 & 0.0000 & 0.0000 & 0.0030 & 0.0016 \\
    Tonga & 1590  & 7     & 0.0000 & 0.0000 & 0.0000 & 0.0000 & 0.0010 & 0.0012 \\
    \hdashline
    Italy & 152152 & 8     & 0.0803 & 0.0500 & 0.0227 & 0.0119 & 0.0241 & 0.0202 \\
    Greece & 25741 & 8     & 0.0008 & 0.0011 & 0.0209 & 0.0119 & 0.0241 & 0.0159 \\
    Portugal & 22053 & 8     & 0.0008 & 0.0000 & 0.0164 & 0.0119 & 0.0241 & 0.0129 \\
    Malta & 3135  & 8     & 0.0000 & 0.0000 & 0.0000 & 0.0000 & 0.0046 & 0.0037 \\
    Albania & 2845  & 8     & 0.0000 & 0.0000 & 0.0000 & 0.0000 & 0.0046 & 0.0037 \\
    San Marino & 1944  & 8     & 0.0000 & 0.0000 & 0.0000 & 0.0000 & 0.0019 & 0.0024 \\
    \hdashline
    France & 203003 & 9     & 0.1079 & 0.0500 & 0.0227 & 0.0357 & 0.0833 & 0.0588 \\
    \hdashline
    United Kingdom & 203003 & 10    & 0.1079 & 0.0500 & 0.0227 & 0.0357 & 0.0833 & 0.0588 \\
    \hdashline
    Korea & 87279 & 11    & 0.0101 & 0.0500 & 0.0227 & 0.0000 & 0.0145 & 0.0135 \\
    Australia & 67176 & 11    & 0.0061 & 0.0495 & 0.0227 & 0.0000 & 0.0145 & 0.0135 \\
    New Zealand & 13973 & 11    & 0.0008 & 0.0000 & 0.0025 & 0.0000 & 0.0145 & 0.0080 \\
    Mongolia & 2175  & 11    & 0.0000 & 0.0000 & 0.0000 & 0.0000 & 0.0029 & 0.0031 \\
    Papua New Guinea & 2084  & 11    & 0.0000 & 0.0000 & 0.0000 & 0.0000 & 0.0029 & 0.0029 \\
    Vanuatu & 1690  & 11    & 0.0000 & 0.0000 & 0.0000 & 0.0000 & 0.0022 & 0.0018 \\
    Seychelles & 1681  & 11    & 0.0000 & 0.0000 & 0.0000 & 0.0000 & 0.0022 & 0.0018 \\
    Solomon Islands & 1660  & 11    & 0.0000 & 0.0000 & 0.0000 & 0.0000 & 0.0021 & 0.0017 \\
    Samoa & 1614  & 11    & 0.0000 & 0.0000 & 0.0000 & 0.0000 & 0.0021 & 0.0017 \\
    Kiribati & 1564  & 11    & 0.0000 & 0.0000 & 0.0000 & 0.0000 & 0.0020 & 0.0016 \\
    Micronesia, Federated States of & 1524  & 11    & 0.0000 & 0.0000 & 0.0000 & 0.0000 & 0.0020 & 0.0015 \\
    Marshall Islands & 1501  & 11    & 0.0000 & 0.0000 & 0.0000 & 0.0000 & 0.0019 & 0.0015 \\
    Palau & 1501  & 11    & 0.0000 & 0.0000 & 0.0000 & 0.0000 & 0.0019 & 0.0015 \\
    Nauru & 1480  & 11    & 0.0000 & 0.0000 & 0.0000 & 0.0000 & 0.0019 & 0.0015 \\
    Tuvalu & 1477  & 11    & 0.0000 & 0.0000 & 0.0000 & 0.0000 & 0.0019 & 0.0015 \\
    \hdashline
     Canada & 111691 & 12    & 0.0223 & 0.0500 & 0.0227 & 0.0000 & 0.0082 & 0.0169 \\
    Ireland & 35951 & 12    & 0.0016 & 0.0053 & 0.0227 & 0.0000 & 0.0082 & 0.0149 \\
    Jamaica & 5281  & 12    & 0.0000 & 0.0000 & 0.0000 & 0.0000 & 0.0013 & 0.0052 \\
    Bahamas, The & 3276  & 12    & 0.0000 & 0.0000 & 0.0000 & 0.0000 & 0.0013 & 0.0032 \\
    Barbados & 2397  & 12    & 0.0000 & 0.0000 & 0.0000 & 0.0000 & 0.0013 & 0.0026 \\
    Belize & 1719  & 12    & 0.0000 & 0.0000 & 0.0000 & 0.0000 & 0.0011 & 0.0021 \\

\bottomrule
\end{tabular}%
\end{adjustbox}
\end{table}

\begin{table}[htbp]
\centering
\caption*{Table 1: Voting Power at the IMF in March 2025 (\textit{Continued})}
\begin{adjustbox}{width=0.95\textwidth, totalheight=\textheight, keepaspectratio}
\begin{tabular}{lcccccccc}
\toprule
\multirow[b]{2}{*}{\textbf{Country}} & \multirow[b]{2}{*}{\makecell{\textbf{Voting}\\\textbf{weight}}} & \multirow[b]{2}{*}{\textbf{Constituency}} &
\multicolumn{3}{c}{\textbf{Felsenthal index}} &
\multicolumn{3}{c}{\textbf{Felsenthal Owen index}} \\
\cmidrule(lr){4-6} \cmidrule(lr){7-9}
& & & $q=50\%$ & $q=70\%$ & $q=85\%$ & $q=50\%$ & $q=70\%$ & $q=85\%$ \\[0.2em]
\midrule
    
    St. Lucia & 1666  & 12    & 0.0000 & 0.0000 & 0.0000 & 0.0000 & 0.0011 & 0.0018 \\
    Antigua and Barbuda & 1652  & 12    & 0.0000 & 0.0000 & 0.0000 & 0.0000 & 0.0011 & 0.0018 \\
    Grenada & 1616  & 12    & 0.0000 & 0.0000 & 0.0000 & 0.0000 & 0.0011 & 0.0016 \\
    St. Kitts and Nevis & 1577  & 12    & 0.0000 & 0.0000 & 0.0000 & 0.0000 & 0.0011 & 0.0016 \\
    St. Vincent and the Grenadines & 1569  & 12    & 0.0000 & 0.0000 & 0.0000 & 0.0000 & 0.0011 & 0.0016 \\
     \hdashline
    Dominica & 1567  & 12    & 0.0000 & 0.0000 & 0.0000 & 0.0000 & 0.0011 & 0.0016 \\
    Sweden & 45752 & 13    & 0.0024 & 0.0121 & 0.0227 & 0.0000 & 0.0052 & 0.0101 \\
    Norway & 38999 & 13    & 0.0020 & 0.0074 & 0.0227 & 0.0000 & 0.0052 & 0.0101 \\
    Denmark & 35846 & 13    & 0.0016 & 0.0053 & 0.0227 & 0.0000 & 0.0052 & 0.0101 \\
    Finland & 25558 & 13    & 0.0008 & 0.0011 & 0.0208 & 0.0000 & 0.0052 & 0.0086 \\
    Lithuania & 5868  & 13    & 0.0000 & 0.0000 & 0.0001 & 0.0000 & 0.0017 & 0.0031 \\
    Latvia & 4775  & 13    & 0.0000 & 0.0000 & 0.0000 & 0.0000 & 0.0017 & 0.0030 \\
    Iceland & 4670  & 13    & 0.0000 & 0.0000 & 0.0000 & 0.0000 & 0.0017 & 0.0030 \\
    Estonia & 3888  & 13    & 0.0000 & 0.0000 & 0.0000 & 0.0000 & 0.0017 & 0.0028 \\
     \hdashline
    Turkey & 48038 & 14    & 0.0028 & 0.0168 & 0.0227 & 0.0000 & 0.0023 & 0.0097 \\
    Austria & 40772 & 14    & 0.0020 & 0.0079 & 0.0227 & 0.0000 & 0.0023 & 0.0097 \\
    Czechia & 23254 & 14    & 0.0008 & 0.0000 & 0.0183 & 0.0000 & 0.0023 & 0.0085 \\
    Hungary & 20852 & 14    & 0.0008 & 0.0000 & 0.0141 & 0.0000 & 0.0023 & 0.0083 \\
    Slovak Republic & 11462 & 14    & 0.0008 & 0.0000 & 0.0010 & 0.0000 & 0.0017 & 0.0056 \\
    Belarus & 8267  & 14    & 0.0004 & 0.0000 & 0.0002 & 0.0000 & 0.0012 & 0.0042 \\
    Slovenia & 7317  & 14    & 0.0004 & 0.0000 & 0.0001 & 0.0000 & 0.0012 & 0.0032 \\
    Kosovo & 2278  & 14    & 0.0000 & 0.0000 & 0.0000 & 0.0000 & 0.0006 & 0.0015 \\
     \hdashline
    Sri Lanka & 132596 & 15    & 0.0527 & 0.0500 & 0.0227 & 0.0000 & 0.0069 & 0.0303 \\
    Bhutan & 12118 & 15    & 0.0008 & 0.0000 & 0.0013 & 0.0000 & 0.0069 & 0.0085 \\
    Bangladesh & 6150  & 15    & 0.0000 & 0.0000 & 0.0001 & 0.0000 & 0.0000 & 0.0057 \\
    India & 1656  & 15    & 0.0000 & 0.0000 & 0.0000 & 0.0000 & 0.0000 & 0.0022 \\
     \hdashline
    Tajikistan & 59163 & 16    & 0.0049 & 0.0411 & 0.0227 & 0.0000 & 0.0015 & 0.0094 \\
    Serbia & 42406 & 16    & 0.0024 & 0.0084 & 0.0227 & 0.0000 & 0.0015 & 0.0094 \\
    Kyrgyz Republic & 13036 & 16    & 0.0008 & 0.0000 & 0.0018 & 0.0000 & 0.0015 & 0.0068 \\
    Switzerland & 8000  & 16    & 0.0004 & 0.0000 & 0.0002 & 0.0000 & 0.0015 & 0.0048 \\
    Azerbaijan & 7240  & 16    & 0.0004 & 0.0000 & 0.0001 & 0.0000 & 0.0015 & 0.0046 \\
    Kazakhstan & 5369  & 16    & 0.0000 & 0.0000 & 0.0000 & 0.0000 & 0.0015 & 0.0039 \\
    Uzbekistan & 3838  & 16    & 0.0000 & 0.0000 & 0.0000 & 0.0000 & 0.0015 & 0.0023 \\
    Liechtenstein & 3228  & 16    & 0.0000 & 0.0000 & 0.0000 & 0.0000 & 0.0010 & 0.0021 \\
    Turkmenistan & 3192  & 16    & 0.0000 & 0.0000 & 0.0000 & 0.0000 & 0.0010 & 0.0020 \\
    Poland & 2452  & 16    & 0.0000 & 0.0000 & 0.0000 & 0.0000 & 0.0010 & 0.0015 \\
    \hdashline
    Syrian Arab Republic & 130489 & 17    & 0.0462 & 0.0500 & 0.0227 & 0.0000 & 0.0000 & 0.0304 \\
    Russian Federation & 6964  & 17    & 0.0004 & 0.0000 & 0.0001 & 0.0000 & 0.0000 & 0.0041 \\
    \hdashline
     Iran, Islamic Republic of & 37123 & 18    & 0.0016 & 0.0058 & 0.0227 & 0.0000 & 0.0000 & 0.0055 \\
    Pakistan & 21762 & 18    & 0.0008 & 0.0000 & 0.0159 & 0.0000 & 0.0000 & 0.0050 \\
    Algeria & 21051 & 18    & 0.0008 & 0.0000 & 0.0145 & 0.0000 & 0.0000 & 0.0050 \\

\bottomrule
\end{tabular}%
\end{adjustbox}
\end{table}

\begin{table}[htbp]
\centering
\caption*{Table 1: Voting Power at the IMF in March 2025 (\textit{Continued})}
\begin{adjustbox}{width=0.95\textwidth, totalheight=\textheight, keepaspectratio}
\begin{tabular}{lcccccccc}
\toprule
\multirow[b]{2}{*}{\textbf{Country}} & \multirow[b]{2}{*}{\makecell{\textbf{Voting}\\\textbf{weight}}} & \multirow[b]{2}{*}{\textbf{Constituency}} &
\multicolumn{3}{c}{\textbf{Felsenthal index}} &
\multicolumn{3}{c}{\textbf{Felsenthal Owen index}} \\
\cmidrule(lr){4-6} \cmidrule(lr){7-9}
& & & $q=50\%$ & $q=70\%$ & $q=85\%$ & $q=50\%$ & $q=70\%$ & $q=85\%$ \\[0.2em]
\midrule

    Libya & 17184 & 18    & 0.0008 & 0.0000 & 0.0065 & 0.0000 & 0.0000 & 0.0048 \\
    Morocco & 10396 & 18    & 0.0008 & 0.0000 & 0.0006 & 0.0000 & 0.0000 & 0.0029 \\
    Ghana & 8832  & 18    & 0.0008 & 0.0000 & 0.0003 & 0.0000 & 0.0000 & 0.0027 \\
    Tunisia & 6904  & 18    & 0.0004 & 0.0000 & 0.0001 & 0.0000 & 0.0000 & 0.0024 \\
    \hdashline
    United Arab Emirates & 24564 & 19    & 0.0008 & 0.0011 & 0.0199 & 0.0000 & 0.0000 & 0.0029 \\
    Kuwait & 20787 & 19    & 0.0008 & 0.0000 & 0.0139 & 0.0000 & 0.0000 & 0.0029 \\
    Iraq  & 18090 & 19    & 0.0008 & 0.0000 & 0.0082 & 0.0000 & 0.0000 & 0.0029 \\
    Qatar & 8803  & 19    & 0.0008 & 0.0000 & 0.0003 & 0.0000 & 0.0000 & 0.0025 \\
    Lebanon & 7787  & 19    & 0.0004 & 0.0000 & 0.0002 & 0.0000 & 0.0000 & 0.0023 \\
    Oman  & 6896  & 19    & 0.0004 & 0.0000 & 0.0001 & 0.0000 & 0.0000 & 0.0021 \\
    Yemen, Republic of & 6322  & 19    & 0.0000 & 0.0000 & 0.0001 & 0.0000 & 0.0000 & 0.0018 \\
    Egypt & 5402  & 19    & 0.0000 & 0.0000 & 0.0000 & 0.0000 & 0.0000 & 0.0017 \\
    Jordan & 4883  & 19    & 0.0000 & 0.0000 & 0.0000 & 0.0000 & 0.0000 & 0.0011 \\
    Bahrain & 4388  & 19    & 0.0000 & 0.0000 & 0.0000 & 0.0000 & 0.0000 & 0.0011 \\
    Somalia & 3086  & 19    & 0.0000 & 0.0000 & 0.0000 & 0.0000 & 0.0000 & 0.0007 \\
    Maldives & 1664  & 19    & 0.0000 & 0.0000 & 0.0000 & 0.0000 & 0.0000 & 0.0003 \\
    \hdashline
    Saudi Arabia & 101378 & 20    & 0.0178 & 0.0500 & 0.0227 & 0.0000 & 0.0000 & 0.0142 \\
    \hdashline
    Argentina & 33325 & 21    & 0.0016 & 0.0032 & 0.0227 & 0.0000 & 0.0000 & 0.0005 \\
    Chile & 18895 & 21    & 0.0008 & 0.0000 & 0.0098 & 0.0000 & 0.0000 & 0.0005 \\
    Peru  & 14797 & 21    & 0.0008 & 0.0000 & 0.0033 & 0.0000 & 0.0000 & 0.0005 \\
    Uruguay & 5743  & 21    & 0.0000 & 0.0000 & 0.0001 & 0.0000 & 0.0000 & 0.0002 \\
    Bolivia & 3853  & 21    & 0.0000 & 0.0000 & 0.0000 & 0.0000 & 0.0000 & 0.0002 \\
    Paraguay & 3466  & 21    & 0.0000 & 0.0000 & 0.0000 & 0.0000 & 0.0000 & 0.0002 \\
    \hdashline
    Nigeria & 25997 & 22    & 0.0008 & 0.0011 & 0.0211 & 0.0000 & 0.0000 & 0.0001 \\
     Côte d'Ivoire & 7956  & 22    & 0.0004 & 0.0000 & 0.0002 & 0.0000 & 0.0000 & 0.0001 \\
    Senegal & 4688  & 22    & 0.0000 & 0.0000 & 0.0000 & 0.0000 & 0.0000 & 0.0001 \\
    Liberia & 4036  & 22    & 0.0000 & 0.0000 & 0.0000 & 0.0000 & 0.0000 & 0.0001 \\
    Guinea & 3594  & 22    & 0.0000 & 0.0000 & 0.0000 & 0.0000 & 0.0000 & 0.0001 \\
    Sierra Leone & 3526  & 22    & 0.0000 & 0.0000 & 0.0000 & 0.0000 & 0.0000 & 0.0001 \\
    Mali  & 3318  & 22    & 0.0000 & 0.0000 & 0.0000 & 0.0000 & 0.0000 & 0.0001 \\
    Togo  & 2920  & 22    & 0.0000 & 0.0000 & 0.0000 & 0.0000 & 0.0000 & 0.0001 \\
    Niger & 2768  & 22    & 0.0000 & 0.0000 & 0.0000 & 0.0000 & 0.0000 & 0.0001 \\
    Mauritania & 2740  & 22    & 0.0000 & 0.0000 & 0.0000 & 0.0000 & 0.0000 & 0.0001 \\
    Benin & 2690  & 22    & 0.0000 & 0.0000 & 0.0000 & 0.0000 & 0.0000 & 0.0001 \\
    Burkina Faso & 2656  & 22    & 0.0000 & 0.0000 & 0.0000 & 0.0000 & 0.0000 & 0.0001 \\
    Gambia, The & 2074  & 22    & 0.0000 & 0.0000 & 0.0000 & 0.0000 & 0.0000 & 0.0001 \\
    Guinea-Bissau & 1736  & 22    & 0.0000 & 0.0000 & 0.0000 & 0.0000 & 0.0000 & 0.0001 \\
    \hdashline
    Congo, Democratic Republic of the & 12112 & 23    & 0.0008 & 0.0000 & 0.0013 & 0.0000 & 0.0000 & 0.0001 \\
    Sudan & 7754  & 23    & 0.0004 & 0.0000 & 0.0002 & 0.0000 & 0.0000 & 0.0001 \\
    Kenya & 6880  & 23    & 0.0004 & 0.0000 & 0.0001 & 0.0000 & 0.0000 & 0.0001 \\
    Uganda & 5062  & 23    & 0.0000 & 0.0000 & 0.0000 & 0.0000 & 0.0000 & 0.0001 \\
\bottomrule
\end{tabular}%
\end{adjustbox}
\end{table}
\begin{table}[htbp]
\centering
\caption*{Table 1: Voting Power at the IMF in March 2025 (\textit{Continued})}
\begin{adjustbox}{width=0.95\textwidth, totalheight=\textheight, keepaspectratio}
\begin{tabular}{lcccccccc}
\toprule
\multirow[b]{2}{*}{\textbf{Country}} & \multirow[b]{2}{*}{\makecell{\textbf{Voting}\\\textbf{weight}}} & \multirow[b]{2}{*}{\textbf{Constituency}} &
\multicolumn{3}{c}{\textbf{Felsenthal index}} &
\multicolumn{3}{c}{\textbf{Felsenthal Owen index}} \\
\cmidrule(lr){4-6} \cmidrule(lr){7-9}
& & & $q=50\%$ & $q=70\%$ & $q=85\%$ & $q=50\%$ & $q=70\%$ & $q=85\%$ \\[0.2em]
\midrule
   
Ethiopia & 4459  & 23    & 0.0000 & 0.0000 & 0.0000 & 0.0000 & 0.0000 & 0.0001 \\
    Cameroon & 4212  & 23    & 0.0000 & 0.0000 & 0.0000 & 0.0000 & 0.0000 & 0.0001 \\
    South Sudan, Republic of & 3912  & 23    & 0.0000 & 0.0000 & 0.0000 & 0.0000 & 0.0000 & 0.0001 \\
    Gabon & 3612  & 23    & 0.0000 & 0.0000 & 0.0000 & 0.0000 & 0.0000 & 0.0001 \\
    Congo, Republic of & 3072  & 23    & 0.0000 & 0.0000 & 0.0000 & 0.0000 & 0.0000 & 0.0001 \\
    Rwanda & 3054  & 23    & 0.0000 & 0.0000 & 0.0000 & 0.0000 & 0.0000 & 0.0001 \\
    Equatorial Guinea & 3027  & 23    & 0.0000 & 0.0000 & 0.0000 & 0.0000 & 0.0000 & 0.0001 \\
    Burundi & 2992  & 23    & 0.0000 & 0.0000 & 0.0000 & 0.0000 & 0.0000 & 0.0001 \\
    Chad  & 2854  & 23    & 0.0000 & 0.0000 & 0.0000 & 0.0000 & 0.0000 & 0.0001 \\
    Central African Republic & 2566  & 23    & 0.0000 & 0.0000 & 0.0000 & 0.0000 & 0.0000 & 0.0001 \\
    Djibouti & 1770  & 23    & 0.0000 & 0.0000 & 0.0000 & 0.0000 & 0.0000 & 0.0001 \\
    Eritrea & 1611  & 23    & 0.0000 & 0.0000 & 0.0000 & 0.0000 & 0.0000 & 0.0001 \\
    São Tomé and Príncipe & 1600  & 23    & 0.0000 & 0.0000 & 0.0000 & 0.0000 & 0.0000 & 0.0001 \\
    \hdashline
    Zambia & 11234 & 24    & 0.0008 & 0.0000 & 0.0009 & 0.0000 & 0.0000 & 0.0000 \\
    Angola & 8853  & 24    & 0.0008 & 0.0000 & 0.0003 & 0.0000 & 0.0000 & 0.0000 \\
    Zimbabwe & 8520  & 24    & 0.0004 & 0.0000 & 0.0003 & 0.0000 & 0.0000 & 0.0000 \\
    Tanzania & 5430  & 24    & 0.0000 & 0.0000 & 0.0000 & 0.0000 & 0.0000 & 0.0000 \\
    Madagascar & 3896  & 24    & 0.0000 & 0.0000 & 0.0000 & 0.0000 & 0.0000 & 0.0000 \\
    Mozambique & 3724  & 24    & 0.0000 & 0.0000 & 0.0000 & 0.0000 & 0.0000 & 0.0000 \\
    Botswana & 3424  & 24    & 0.0000 & 0.0000 & 0.0000 & 0.0000 & 0.0000 & 0.0000 \\
    Namibia & 3363  & 24    & 0.0000 & 0.0000 & 0.0000 & 0.0000 & 0.0000 & 0.0000 \\
    Mauritius & 2874  & 24    & 0.0000 & 0.0000 & 0.0000 & 0.0000 & 0.0000 & 0.0000 \\
    Malawi & 2840  & 24    & 0.0000 & 0.0000 & 0.0000 & 0.0000 & 0.0000 & 0.0000 \\
    Eswatini & 2237  & 24    & 0.0000 & 0.0000 & 0.0000 & 0.0000 & 0.0000 & 0.0000 \\
    Lesotho & 2150  & 24    & 0.0000 & 0.0000 & 0.0000 & 0.0000 & 0.0000 & 0.0000 \\
    Comoros & 1630  & 24    & 0.0000 & 0.0000 & 0.0000 & 0.0000 & 0.0000 & 0.0000 \\
    South Africa & 319   & 24    & 0.0000 & 0.0000 & 0.0000 & 0.0000 & 0.0000 & 0.0000 \\
    \hdashline
    Guyana & 8429  & 25    & 0.0004 & 0.0000 & 0.0003 & 0.0000 & 0.0000 & 0.0000 \\
    Ecuador & 6226  & 25    & 0.0000 & 0.0000 & 0.0001 & 0.0000 & 0.0000 & 0.0000 \\
    Suriname & 5220  & 25    & 0.0000 & 0.0000 & 0.0000 & 0.0000 & 0.0000 & 0.0000 \\
    Panama & 4052  & 25    & 0.0000 & 0.0000 & 0.0000 & 0.0000 & 0.0000 & 0.0000 \\
    Haiti & 3270  & 25    & 0.0000 & 0.0000 & 0.0000 & 0.0000 & 0.0000 & 0.0000 \\
    Nicaragua & 3090  & 25    & 0.0000 & 0.0000 & 0.0000 & 0.0000 & 0.0000 & 0.0000 \\
    Timor-Leste & 2741  & 25    & 0.0000 & 0.0000 & 0.0000 & 0.0000 & 0.0000 & 0.0000 \\
    Trinidad and Tobago & 1708  & 25    & 0.0000 & 0.0000 & 0.0000 & 0.0000 & 0.0000 & 0.0000 \\
    Dominican Republic & 1689  & 25    & 0.0000 & 0.0000 & 0.0000 & 0.0000 & 0.0000 & 0.0000 \\
    Cabo Verde & 872   & 25    & 0.0000 & 0.0000 & 0.0000 & 0.0000 & 0.0000 & 0.0000 \\
    Brazil & 111   & 25    & 0.0000 & 0.0000 & 0.0000 & 0.0000 & 0.0000 & 0.0000 \\

\bottomrule
\end{tabular}%
\end{adjustbox}
\end{table}